\documentclass[10pt,twoside,openright]{report}
\usepackage{epsfig}
\usepackage{latexsym,supertabular}

\begin{document}
\newcommand{\DS}{\displaystyle}
\renewcommand{\topfraction}{0.85}
\renewcommand{\textfraction}{0.15}
\setcounter{page}{1}
\thispagestyle{empty}
\begin{center}
\vspace*{\fill}
\textbf{\Huge Coulomb blockade effects\\
in anodised\\
niobium nanostructures\\~}\par
\vspace*{8mm}
\textbf{\Large (Coulomb-blockad-effekter\\
i anodiserade Nb-nanostrukturer)\\~}
\vfill\vfill
{\LARGE Dipl.-Phys. Torsten Henning}
\vfill
{\Large Licentiate thesis/Licentiatuppsats}
\vskip24mm
\epsfig{file=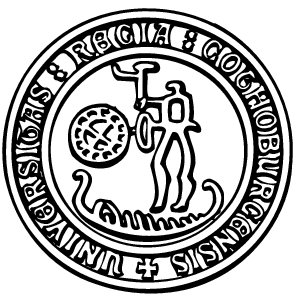,height=30mm}\hspace{12mm}
\epsfig{file=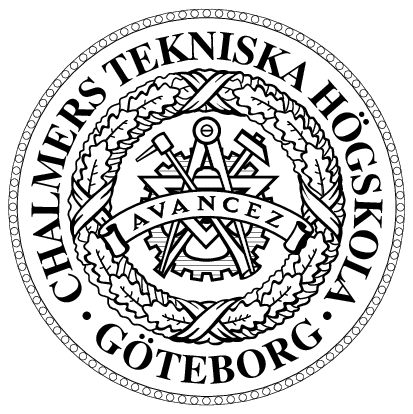,height=30mm}
\vskip8mm
Institutionen f\"or Mikroelektronik och Nanovetenskap\\
Chalmers Tekniska H\"ogskola AB och G\"oteborgs Universitet\\
1997-05-14
\end{center}
\vspace*{8mm} 
\clearpage
\setcounter{page}{2}\thispagestyle{empty}
\thispagestyle{empty}
\vspace*{\fill}
\begin{center}
WWW edition, 1997-10-03\\
\copyright{} This copy of the thesis has been supplied on condition that
anyone who consults it is understood to recognise that its copyright
rests with the author and that no quotation from the thesis, nor any
information derived therefrom, may be published without the author's
prior written consent.
\end{center}
\newpage
\setcounter{page}{3}
\thispagestyle{plain}
\vspace*{\fill}
\centerline{{\bf Abstract}}
\vspace{3ex}
Niobium thin film nanostructures were fabricated using electron beam
lithography and a liftoff process involving a four layer resist system. 
Wires having micrometres in length and below 200\,nm in width were
deposited on an insulating substrate and subsequently thinned by
anodisation. 
The resistance of the wires was monitored in situ and could be trimmed
by controlling the anodisation voltage and time.
The method appears appropriate for the fabrication of very small
resistors for use as biasing  resistors for single
electron devices.
Transport properties of these resistors were measured at millikelvin
temperatures. 
They had nonlinear current-voltage characteristics around zero bias and
showed a transistion from superconducting to insulating behaviour.
Analysis of the offset voltage showed that a Coulomb blockade set in
when the sheet resistance of the films exceeded the superconducting
quantum resistance $\approx 6.45$\,k$\Omega$.\par
Samples in a single electron transistor-like geometry with two variable
thickness weak links were made by combined angular evaporation of
niobium and anodisation. Overlapping gate electrodes were deposited on
these samples. The drain-source current-voltage characteristics could be
modulated by the gate voltage, giving a response typical for a system of
multiple ultrasmall tunnel junctions.
\vfill\vfill
\vfill\newpage
\setcounter{page}{4}
\thispagestyle{plain}
\vspace*{\fill}
\centerline{{\bf Sammanfattning}}
\vspace{3ex}
Tunnfilms-nanostrukturer av niob tillverkades med hj\"alp av
elektron\-str\aa{}le\-lito\-grafi 
och en liftoff-process som anv\"ander sig av
ett fyrlagers-resistsystem. 
Ledningar av n\aa{}gra mikrometers l\"angd och
en bredd mindre \"an 200\,nm deponerades p\aa{} ett isolerande substrat och
f\"ortunnades sedan genom anodisering. Ledningarnas resistans m\"attes
kontinuerligt under anodiseringen
och kunde trimmas genom att kontrollera
anodiseringssp\"anningen och -tiden.
Metoden f\"orefaller l\"ampad f\"or tillverkning av mycket sm\aa{}
resistorer som beh\"ovs som biaseringsresistorer
f\"or en-elek\-tron-kom\-po\-nen\-ter.
Transportegenskaperna hos resistorerna m\"attes vid
milli\-kelvin-tem\-pera\-turer, som visade ickelinj\"ara
str\"om-sp\"annings-karak\-teristika vid l\aa{}ga sp\"anningar
och en
\"overg\aa{}ng fr\aa{}n supraledande till isolerande uppf\"orande.
Analys av f\"or\-skjut\-nings\-sp\"an\-ningen gav att en 
Cou\-lomb\--blockad 
fram\-tr\"adde
n\"ar filmernas ytmotst\aa{}nd \"overskred den supraledande
kvantresistansen $\approx 6.45$\,k$\Omega$. \par
Prover liknande  enkelelektron-transistorer med tv\aa{} svaga l\"ankar
skapades genom kombinerad skuggf\"or\aa{}ngning av niob och
anodisering. En \"overlappande 
grind\-elek\-trod deponerades p\aa{} dessa
prover. Provernas 
str\"om-sp\"an\-nings-karak\-ter\-istika kunde
p\aa{}\-verkas av grindsp\"anningen, och de resulterande kurvorna
liknar dem som f\aa{}s i system av m\aa{}nga ultrasm\aa{}
tunnel\"overg\aa{}ngar. 
\vfill\vfill
\paragraph{Nyckelord:}
niob, anodisering, nanofabrikation,
Coulomb-blockad, supra\-le\-dare-iso\-lator-\"overg\aa{}ng
\vfill
\newpage
\thispagestyle{plain}\tableofcontents
\cleardoublepage\pagestyle{myheadings}\markboth{}{}
\addcontentsline{toc}{chapter}{List of Figures}\listoffigures
\cleardoublepage\pagestyle{myheadings}\markboth{}{}
\addcontentsline{toc}{chapter}{Preface}
\chapter*{Preface}
\begin{flushright}
\textsl{`I have thought about some of the problems\\
of building electric circuits on a small scale,\\
and the problem of resistance is serious.'}\\
(Richard P. Feynman, 1959\cite{feynman:60:plenty})
\end{flushright}
\par
\vspace*{3ex}

Before we start with any scientific stuff, let me thank some people who
have contributed a lot to this thesis, or to the fact that I
could write it here. First of all Tord Claeson, who gave me the chance
to start working in his group. Then of course Bengt Nilsson, for many
good suggestions and most of all for keeping the SnL up and running so
smoothly that we hardly know how to spell dauntajm. 
Thanks to all those who
keep paperwork from us and make the SnL probably the least bureaucratic
nanofabrication facility in the world, especially Ann-Marie
Frykestig. Thanks to Staffan Pehrson and Henrik Frederiksen for many
small and some really big things they built and serviced. 
I'm mainly thinking about the Niobium System, of course. Peter
Wahlgren also invested a lot of time into that machine.
Last but not
least Per Delsing and David Haviland, who initiated and supervised this
work, and the numerous PhD students, past and present, who helped in one
way or another.\par
A more formal, but nonetheless
cordial thankyou to the German Academic Exchange
Service DAAD, who financed my first year here `im Rahmen des Zweiten
Hochschulsonderprogramms', and to the Commission of the European
Communities, who support me presently. This work was part of ESPRIT
programme 9005 SETTRON. \par
And let's not forget the taxpayers in Germany, Sweden and Europe in
general who once earned the money.\par

Partial results have been published  in  conference
proceedings:
\begin{enumerate}
\item Torsten Henning, D.~B. Haviland, P.~Delsing: 
  \textit{Transition from supercurrent to Coulomb blockade tuned by 
   anodization of Nb wires.}
   Czechoslovak Journal of Physics, Vol. 46 (1996), Suppl. S4, pp. 2341-2342
\item Torsten Henning, D. B. Haviland, P. Delsing: 
   \textit{Fabrication of Coulomb blockade elements with an electrolytic
   anodization process.}
   Symposium on Single-Electron nanoelectronics, 
  190th Meeting of the Electrochemical Society (San Antonio, 6--11
  October 1996)
\item Torsten Henning, D. B. Haviland, P. Delsing: 
   \textit{Charging effects and superconductivity
   in anodised niobium nanostructures.}
   6th International Superconductive Electronics Conference
   (Berlin, 25--28~June 1997)
\end{enumerate}
and at the Spring Meetings of the German Physical Society. 
A paper has been submitted:
\begin{itemize}
\item[4] Torsten Henning, D.~B. Haviland, P.~Delsing:
   \textit{Coulomb blockade effects in anodised niobium 
   nanostructures.}
   submitted to Superconductor Science and Technology 1997-05-09
\end{itemize}

The International system of Units is used exclusively throughout this
work. In questions of style connected with the SI, I follow largely the
recommendations of the National Institute of Standards and Technology
\cite{taylor:95:si}, not so much because I think America is leading in
the sound use of units, but because they were so kind to publish their
booklet free of charge on the WWW. Appendix~\ref{symbolsapp} explains
the nomenclature used.
Making graphic representations of data is a science in itself (and
not a question of taste). I recommend the
book by Cleveland \cite{cleveland:85:graph}, whose guidelines I
tried to follow.

Nanofabrication, like all branches of technology, has developed its own
set of acronyms and abbreviations. 
I have  collected as much jargon as possible in
a glossary that you will find as appendix \ref{glossary}. 

Appendix \ref{recipes} lists in detail all the processes that were
developed or used in the course of this work. 
Recipes are 
always subject to change, and the appendix reflects the
latest state of the art, so samples described in this thesis may have
been fabricated by a process with different parameters.

\paragraph{Appendix omitted from the WWW edition:}
This thesis is concluded by a seemingly unrelated appendix on the
lateral nanostructuring of II-VI semiconductor heterostructures. 
Well, not so unrelated, because there is a lot of general
nanofabrication know-how involved that is common to the niobium
nanostructures treated in the main part, and to the quantum dots
mentioned in this appendix. This project was initiated and is
coordinated by Peter Klar, PhD by now, of UEA Norwich. Thanks to him for
giving me the opportunity to do even more useful things with all the
nice equipment here. 
And once again to Bengt Nilsson for working
countless overtime hours and repairing the JEOL just when we needed it
most.
\pagestyle{headings}
\chapter{Introduction, background}
This chapter is intended to give a brief introduction to some of the
basic phenomena and concepts relevant for the work covered in this
report. 
For a more thorough introduction, there is a
lot of reference material and textbooks on tunnelling, 
superconductivity \cite{tinkham:80:intro}, and
charging effects 
\cite{schoen:90:review,altshuler:91:mpsbook,grabert:92:sctbook}. 
Reference
material on niobium and its anodisation is quoted in the respective
sections below. 
\section{Tunnelling and superconductivity} 
Tunnelling and superconductivity are two quantum phenomena that are
observed in transport measurements. 
\subsection{Tunnelling} 
Tunnelling is a transport mechanism that can only be understood in terms
of quantum mechanics. Consider a conduction electron with energy $E$ near the
Fermi surface in a region that we call `left electrode', and an adjacent
region called `barrier' where the conduction band edge lies at a higher
energy. On the other side of the barrier we have the `right
electrode'. Even if the Fermi energy in the right electrode is 
different from that in the left electrode, classical physics  would
prohibit the transmission of the electron. Quantum mechanics, however,
predicts a nonvanishing transmission through the barrier as long as it is
finite in height (energetic) and width (in real space). \par
A  voltage applied between left and right electrode results in a
current flow. For small voltages, the
current-voltage characteristics are linear and define a tunnelling
resistance $R_T$.
Tunnelling processes are classified as elastic, if the energy of the
tunnelling particle is conserved, or inelastic. In the latter case,
dissipation occurs through excitations in the barrier, the electrodes,
or the electrode-barrier interfaces. 
\begin{figure} 
\begin{center}
\epsfig{file=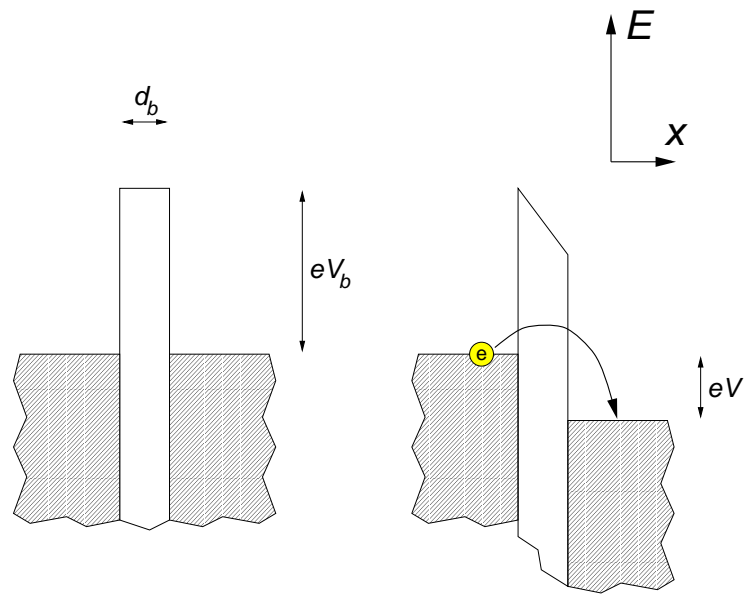,width=0.8\textwidth}
\end{center}
\caption[Tunnelling between metallic conductors]{%
\label{tunnprincfig}Schematic representation of the tunnelling between
two metallic conductors separated by a barrier of height $eV_b$ and
width $d_b$. In the right picture, the junction is biased with a voltage
$V$, favouring (inelastic) tunnelling of electrons
from the left to the right
electrode.}
\end{figure} 
Figure~\ref{tunnprincfig} is a
schematic representation of the tunnelling between two metallic
conductors.
\par
Two examples of systems showing tunnelling are metallic tips showing
field emission, and  structures with a thin insulator separating
two conductors. In the first case, the `left electrode' is the metal tip,
and the barrier is created by the work function, i.\,e. the energy needed
to move an electron from the conductor to infinity. An electric field
tilts the vacuum level, resulting in a triangular shaped barrier that
allows the emission of electrons
into vacuum (`right electrode'). This effect is not only of historical
interest as an early verification of quantum mechanics (observed by
Lilienfeld in 1922  and explained by Fowler and
Nordheim in 1928), it is also of technological
importance as the working principle of advanced electron guns in
electron beam lithography machines. The second example,
conductor-insulator-conductor structures in different varieties, is the
main subject of this report. \par
\subsection{Superconductivity} 
Below a critical temperature $T_c$ and critical magnetic field $H_c$,
certain materials are in a thermodynamic state called the
`superconducting state'. It manifests itself in a number of effects.
The first effect discovered was the vanishing of the electrical
resistance, which gave the phenomenon its name
\cite{royalsociety:35:disc}.
Of at least equal importance is the fact that superconductors expel
magnetic fields, the Meissner-Ochsenfeld effect
\cite{meissner:33:nw}.\par 
We will in this report only be dealing with so-called `low temperature'
superconductors. These materials are rather well understood
theoretically, much better than the `high temperature' superconductors
that were discovered just more than a decade ago
\cite{bednorz:87:epl}. A microscopic theory that is
well confirmed experimentally is that of Bardeen, Cooper and Scrieffer
(BCS, \cite{claeson:74:supcon}). Pairs of electrons in time reversed
states (opposite spin and wave vector) interact by exchange of virtual
phonons, leading to an attractive interaction and the formation of
so-called `Cooper pairs'. A collective effect results in the formation
of a ground state that has a lower energy than the Fermi sphere and a
gap in the density of states around the Fermi energy. The energetic
width of this gap is $2\Delta$, the energy required to break up a Cooper
pair and create an excitation. \par
The density of states
has a singularity at the gap edges.
It was measured by Giaever in tunnelling experiments 
that are a direct verification of the BCS theory.\par
Characteristic material parameters are the gap 
(at zero temperature) $\Delta(0)$, the London
penetration depth $\lambda$, the characteristic length over which a
magnetic field drops at the superconductor surface, and the coherence
length $\xi$, over which the Cooper pair density varies. The critical
temperature depends on the gap at zero temperature, and the gap energy
vanishes at the critical temperature. Niobium is the element with the
highest critical temperature
(at ambient pressure), $T_c\approx 9.2\,$K in bulk, corresponding
to a gap of $2\Delta\approx3$\,mV. 
The gap is reduced in thin films, considerably below 50\,nm thickness
\cite{lehnert:94:nbgap}. 
The London
penetration depth of niobium
is $\lambda=32$\,nm \cite{auer:73:prb}, 
and the coherence
length $\xi=$39\,nm \cite{auer:73:prb}.\par
\subsection{Josephson effects} 
The superconducting condensate is described by a multiparticle wave
function with a phase $\phi_i(x,t)$. A tunnel junction with an oxide barrier 
and superconducting electrodes is
one example of a system where two superconducting condensates are more
or less strongly coupled. For such systems, Josephson
\cite{josephson:64:rmp} predicted several effects that now summarily
bear his name. Important for us is the so-called DC Josephson effect.
Between the two superconductors, a supercurrent can flow, that is a
current without a voltage drop. The maximally possible supercurrent 
(Josephson pair current) has
the following dependence on the phase difference 
$\gamma=\phi_r-\phi_l$
\cite{tinkham:80:intro}:
\begin{equation}
I_c=I_{\it c0}\,\sin\gamma.
\end{equation}
The coupling between the superconductors is often expressed in terms of
a coupling energy
\begin{equation}
E_J=\frac{\DS \hbar}{\DS 2e}\,I_{\it c0}.
\end{equation}
\section{Charging effects} 
In this section, we will give a brief introduction to the subject of
charging energy, proceeding from simple circuits to granular films
rather than historically vice versa.
\subsection{Charging effects in very small tunnel junctions} 
A metal-insulator-metal tunnel junction has a capacitance $C$ that
relates its charge $Q$ to the potential drop $V$ between the
electrodes. Not only does the concept of capacitance apply on the
submicron length scale, even the approximation as a parallel plate
capacitor with capacitance
\begin{equation}
C=\frac{\displaystyle \varepsilon_0\varepsilon A}%
   {\displaystyle d},
\end{equation}
with $d$ the thickness of the oxide barrier,
works quite well for junctions in the submicrometre size
range. \par
\begin{figure} 
\begin{center}
\begin{minipage}[b]{0.6\textwidth}
\caption[Symbol for ultrasmall junction]{\label{jctnsymbfig}Symbol for
an ultrasmall tunnel junction
(bottom), a combination of the symbols for a tunnelling resistance $R_T$
and a capacitance $C$ in parallel (top).}
\end{minipage}\begin{minipage}[b]{0.4\textwidth}
\centering
\epsfig{file=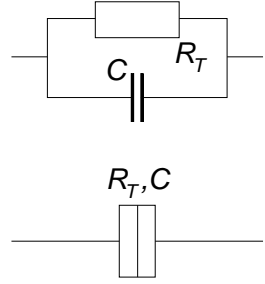,width=0.7\textwidth}
\end{minipage}
\end{center}
\end{figure} 
The symbol used for ultrasmall tunnel junctions (see
figure~\ref{jctnsymbfig}) represents a combination of a tunnelling
resistance $R_T$ and a capacitance $C$. 
\begin{figure} 
\begin{center}
\begin{minipage}[b]{0.5\textwidth}
\caption[Overlap junction]{\label{overlapfig}Implementation of an
ultrasmall junction as an overlapping junction. The barrier is either
formed by oxidising the bottom electrode, e.\,g. in the case of Al as
electrode material, or by depositing a different material and oxidising
it, if applicable.}
\end{minipage}\begin{minipage}[b]{0.5\textwidth}
\centering
\epsfig{file=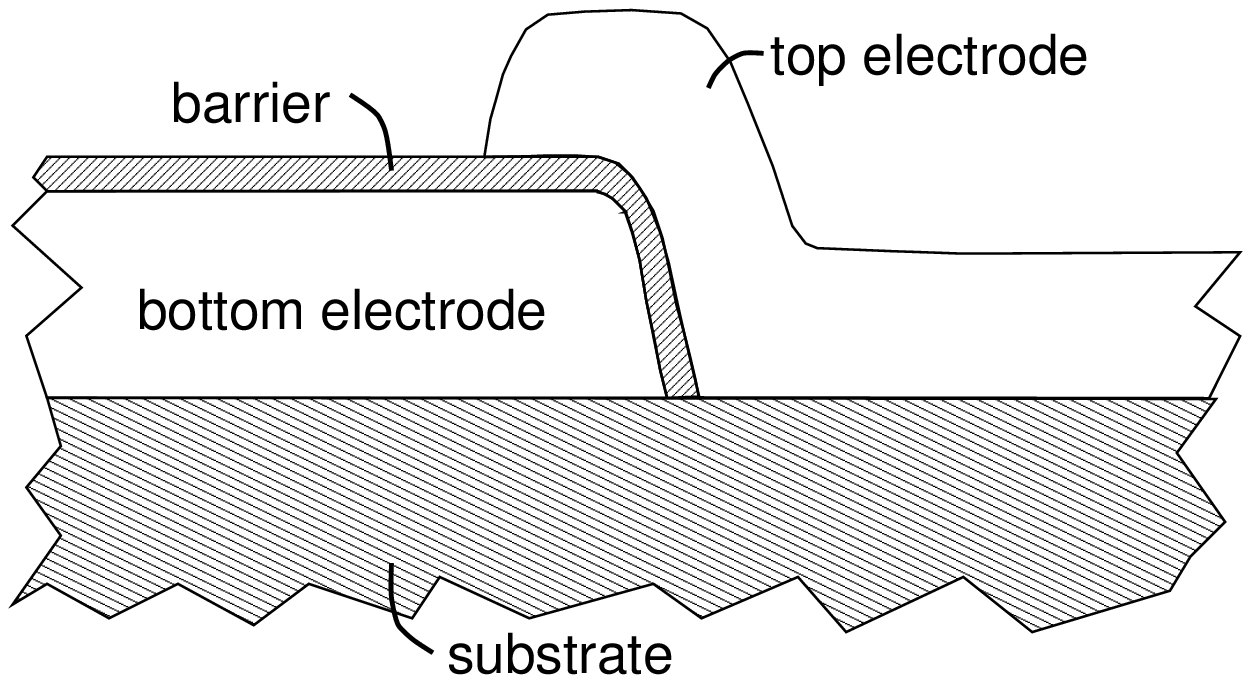,width=0.9\textwidth}
\end{minipage}
\end{center}
\end{figure} 
Figure~\ref{overlapfig}  shows how such a
junction is generally implemented as an overlap junction.\par
Charging the junction to $Q$ requires an energy 
\begin{equation}
E_{\it ch}=\frac{\displaystyle Q^2}{\displaystyle 2C}.
\end{equation}
The charge $Q$ of a junction is a continous variable, since the
(continuous) voltage $V$ can induce any polarisation charge. 
If the charge carriers are localised on either side of the barrier,
tunnelling causes changes of the charge in integer multiples of the
elementary charge $e$. From an argumentation that the characteristic
time of charge leakage though the barrier $R_TC$ should be larger than the
time associated with the energy change during tunnelling via an
uncertainty relation $\Delta\tau\,\Delta E=\hbar/2$, one arrives at a
criterion: charges are localised on either side of the barrier if the
tunnelling resistance $R_T$ is large compared to the quantum resistance
\begin{equation}
R_K=\frac{\displaystyle h}{\displaystyle e^2}.
\end{equation}
The index on $R_K$ expresses the current belief 
\cite{goebel:97:phbl}
that the fraction on the
right is exactly the resistance constant that gives the position of
plateaux in the quantum Hall or Klitzing effect. This
effect is currently being used to realise the resistance standard, and
the Klitzing constant has been defined to 
\begin{equation}
R_{\mbox{\footnotesize K-90}}\equiv 25812.807\,\Omega.
\end{equation}\par
The characteristic charging energy of a junction is that caused by a
single elementary charge,
\begin{equation}
\label{charcharform}
E_C=\frac{\displaystyle e^2}{\displaystyle 2C}.
\end{equation}
For a junction with an area of (100\,nm)$^2$ and a 1\,nm thick oxide
barrier with a dielectric constant of 10, the charging energy
corresponds to a temperature of about 1\,K. To observe single charge
charging effects, the temperature must be lower. Therefore, single
electronics, the science of phenomena related to the charging energy and
the sometimes continuous and sometimes discrete nature of charge, is
today a branch of low temperature physics.\par
The basis of single electronics is the Coulomb blockade. 
If an electron tunnels through a junction, its energy changes by
$\Delta E=E(Q-e)-E(Q)$. No tunnelling occurs unless $\Delta E$ is
negative. 
In an ideal case, this would lead to a
linear current-voltage characteristic that is shifted by 
an offset voltage 
\begin{equation}
V_{\it off}=(E_C/e)(\left|V\right|/V)
\end{equation}
against an Ohmic characteristic.\par
For a junction with superconducting electrodes, the charge of the Cooper
pair $2e$ has to be substituted for $e$ in (\ref{charcharform}).\par
Kulik and Shekhter calculated in 1975 \cite{kulik:75:gran}
the current-voltage characteristics for tunnelling through a small
grain, something that we nowadays would call a double tunnel
junction, with a Coulomb staircase in the asymmetric case. 
In 1982, Widom et al. \cite{widom:82:duality} pointed out the duality
between the Josephson effect and the `current bias frequency effect' that
we now know as `Bloch oscillations'. Likharev and Zorin worked out the
theory of these oscillations in 1985 \cite{likharev:85:jltp}. 
Ben-Jacob et al. predicted similar oscillations of the voltage in
normalconducting current biased junctions \cite{benjacob:85:osc}. 
Soon thereafter, Averin and Likharev published a theory of these
SET-oscillations \cite{averin:86:osc}. A three terminal device based on
the Coulomb blockade and as a dual analogue to the SQUID was introduced
by Likharev as `Single electron transistor' in 1987
\cite{likharev:87:squidieee}. 
\label{cbexppioneers}
The same year, single electron effects
were found experimentally in granular tunnel junctions by Kuzmin and
Likharev \cite{kuzmin:87:jetp} and in lithographically made junctions by
Fulton and Dolan \cite{fulton:87:charprl}.\par
\subsection{Charging effects in granular films} 
Gorter suggested in 1951 \cite{gorter:51:filmres} that the observed
increase of the resistance of thin films at low temperature and low bias
might be due to a granular structure of these films and to charging
effects impeding the charge transfer  between the grains. Neugebauer and
Webb \cite{neugebauer:62:films} treated thin films as a planar array of
small islands, between which charge transfer occurred by
tunnelling. Taking the charging effects into account, they predicted and
measured an Arrhenius type dependence of resistance on temperature (see
page \pageref{myarrh}). The granularity of their samples was
demonstrated by TEM imaging.\par
Giaever and Zeller \cite{giaever:68:part,zeller:69:tunn} made
Al-AlO$_x$-Al junctions with Sn particles embedded in the oxide. They
found effects of the finite size of the grains, causing a finite spacing
of energy levels in the small particle and making it in principle
impossible to align Fermi levels in the electrodes and the
grains. Assuming a distribution of grain sizes could account for the
observed nonlinear current-voltage characteristics, showing what we
today call the Coulomb blockade. Since Sn is a superconductor with a
critical temperature of 3.7\,K (in bulk), above that of Al (1.2\,K),
Giaever and Zeller could also investigate the superconductivity of very
small particles, and claim to have
found that superconductivity persisted down to
the smallest investiagted grain sizes of about 5\,nm. \par
Another experiment with small particles was made by Lambe and Jaklevic
\cite{lambe:69:tunn,jaklevic:75:prb}: they prepared small particles
separated from one electrode by a thin oxide barrier and from the other
electrode by a thick barrier, not transparent for tunnelling electrons
and just providing a capacitive coupling to the grains. In modern terms
we would call this a parallel coupling of different single electron
boxes. They found by capacitance measurements that the grains were
charged stepwise with increasing voltage by tunnelling through the
thinner barrier, and observed a quantum field effect when they modulated
the energy levels in the grains by applying an external electrical
field.
A quantum mechanical treatment with a charging and a tunnelling
Hamiltonian was presented by Kulik and Shekhter \cite{kulik:75:gran} 
in 1975.
\section{Three terminal charging effect devices} 
\begin{figure} 
\begin{center}
\begin{minipage}[b]{0.6\textwidth}
\caption[Single electron transistor]{%
\label{setschemfig}Single electron transistor, biased symmetrically with
respect to ground.}
\end{minipage}\begin{minipage}[b]{0.4\textwidth}
\centering
\epsfig{file=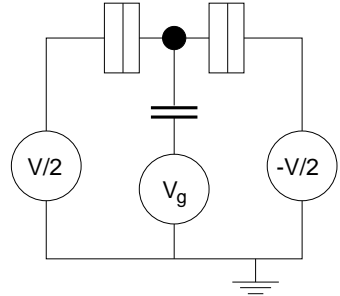,width=0.9\textwidth}
\end{minipage}
\end{center}
\end{figure}

The simples three terminal device based on the charging effect is the
single electron transistor (SET) 
\cite{likharev:87:squidieee},
sketched in fig.~\ref{setschemfig}.
It consists of two ultrasmall junctions in series, isolating an island
from the rest of the circuit. The island is capacitively coupled to a
gate electrode.\par
The electrodes and/or the island can be superconducting. In this case,
the charge carriers are Cooper pairs 
(unless $\Delta \ll E_C$), as well as electrons, and one
speaks of a `superconducting SET' or S-SET rather than using the 
term `charge-effect transistor' (CHET) \cite{amman:89:chet}. \par
The charge of the island is the sum of the charge induced by the gate
voltage 
\begin{equation}
Q_g=C V_g,
\end{equation}
and the number of electrons $n$ in excess of the number of positive charges
(protons) in the island. $Q_g$ is a continuous variable, and the
charging energy $E_{\it ch}$ is a function of $Q_g$ and $n$. For a
derivation of the $I$-$V$ characteristics, see e.\,g. the review article
by
Sch\"on \cite{schoen:96:setchapterpreprint}. Here we will only state
that the ideal SET
\begin{itemize}
\item has a Coulomb blockade whose maximum thresholds are determined by
the junction capacitances, and
\item whose extension is periodic in the gate voltage, with one period
corresponding to one elementary charge induced on the gate capacitance.
\end{itemize}
Ideal means among others that the junctions should be identical;
otherwise, the IVC will show a periodic structure known as the Coulomb
staircase \cite{kulik:75:gran}. 
\section{The need for smaller resistors} 
To understand why 
(physically) very small resistors 
with high resistance are desirable for use in single
electronics, we shall take a brief look at the influence of the
electromagnetic environment on tunnelling in ultrasmall junctions.
\subsection{Coulomb blockade and electromagnetic environment} 
Soon after the first single elctron devices had been made it was
realised that the electromagnetic environment had a profound influence
on their properties 
\cite{nazarov:89:isoljetp,delsing:89:sepprl,cleland:90:fluctuprl,%
devoret:90:envprl}.
The Coulomb blockade in single junction is largely shunted out by the
effect of stray capacitances of the leads. Essentially, these lead
capacitances make the junction voltage 
biased rather than current biased. The
first observations of the Coulomb blockade were therefore made in systems
of two tunnel junctions 
connected in series
(see \ref{cbexppioneers}, where a particular junction
was shielded from the environment by another junction with resistance
higher than the quantum resistance $R_K$.\par
The effect of the electromagnetic environment on ultrasmall tunnel
junctions can be calculated using different approaches. One such
approach is the theory  known
as the `$P(E)$ theory', because it involves a probability density
function $P(E)$ describing the emission or absorption of energy during a
tunnelling process due to the coupling to the environment
\cite{ingold:92:sct,schoen:96:setchapterpreprint}. 
The circuit with the junction(s)
and the environment, represented by an impedance $Z(\omega)$, is
modelled as an ensemble of harmonic oscillators. The approach follows
the model for dissipative quantum mechanics developed by Caldeira and
Leggett \cite{caldeira:83:dissip}.\par
A simple alternative model is the horizon picture 
\cite{nazarov:89:ivcjetp,geigenmueller:89:arrayepl},
in which the electromagnetic environment is treated as purely
capacitive, and its spacial extension as given by the speed of light and
the time corresponding to the energy $eV$ via a time-energy uncertainty
relation. Experiments showed
\cite{delsing:89:sepprl} that it is this time rather than the
traversal time \cite{buettiker:82:tunneltimeprl} that determines
the horizon.
The horizon picture has been found to give reasonable results
for single and double junction circuits in the normalconducting state
\cite{wahlgren:95:prb,wahlgren:95:lic}.\par
In any case, the Coulomb blockade of single junctions is much sharper if
the junction is placed in an environment with high impedance $Z(\omega)$
than in a low-impedance environment. This has been found experimentally
\cite{cleland:92:envprb}, and is rather well understood theoretically
\cite{schoen:96:setchapterpreprint}. Cleland et al. argued that the
environment's impedance can be maximised by contacting the junction with
a high-resistivity material with as small as possible cross section
\cite{cleland:92:envprb}. The impedance of the leads should exceed the
quantum resistance $R_K$ if the Coulomb blockade is not to be
suppressed.\par 
\subsection{Resistors in single electronics} 
The fabrication of such high-resistance low-stray capacitance resistors
is nontrivial. Cleland and coworkers made thin film resistors of NiCr
\cite{cleland:92:envprb} and achieved a `performance' of
30\,k$\Omega/\mu$m. This allowed them to observe that the Coulomb
blockade was much sharper than with leads of 2\,k$\Omega/\mu$m.\par
A series of experiments involving thin film resistors and a single
junction was made by 
Haviland and Kuzmin in
1991\,ff. and led to the observation of the Coulomb blockade of Cooper
pair tunnelling in ultrasmall Josephson junctions and the Bloch
oscillations due to the time correlation caused by the blockade
\cite{haviland:91:zfpb,haviland:91:cpcbepl,kuzmin:91:cpprl,%
kuzmin:91:blochprl,kuzmin:92:blochps}.\par
In all these experiments, the thin film resistors were made from a
different material than the junction electrodes or barriers. This
introduced the need for a third angle evaporation
(using a suspended mask technique), resulting in either a
lot of unwanted material or demanding a complicated tear-off process. In
addition, such thin film resistors have the disadvantage of not being
tunable to a desired resistance value. \par
The motivation for the experiments described in this report was to
develop a fabrication process for resistors that were tunable and
made from the same material one would possibly use as electrode
material, namely niobium. In general, materials with a high sheet
resistance are not easily produced reproducibly. Wong and Ingram
\cite{wong:93:gecu} have reported a germanium-copper alloy for
application in thin film resistors. Still, the difficulty of integrating
this material into the single electronics circuit remains.\par
Two advantages we expect from resistors made by niobium anodisation are
due to the oxide encapsulating the structure. First, the oxide should
protect it against corrosive influence from the atmosphere, and
e.\,g. from chemicals in lithographic processing steps. We found that
indeed our samples, once they `survived' the fabrication process, were
quite stable (with the exception of some of the samples in single
electron transistor-like geometry considered
in~\ref{setlikesection}).\par 
The second expected advantage, which we have not yet verified, is that the
high volume of the resistors should diminish hot electron effects
\cite{wellstood:94:hotelprb}. These effects are caused by the weakness
of the electron-phonon coupling at low temperatures; 
the passage time of the electrons is too short for them to acquire
thermal equilibrium with the lattice (phonons).
Another strategy to
improve cooling and thus reduce hot electron effects is to increase the
volume of the resistor without decreasing its resistance too much
by adding cooling fins \cite{wellstood:94:hotelprb}.\par
An alternative method for making biasing resistors for single electron
devices is the fabrication of arrays of relatively large junctions
\cite{wahlgren:95:lic,pettersson:96:prb}. The integration is
straightforward since these resistors are made in the same angular
evaporation process as the active element(s). The problem of tuning the
resistance is not so serious and mostly a matter of having the right
geometric dimensions since the resistance is determined by the oxide
thickness, and that parameter is critical for the active elements
anyway. As a consequence, for applications not demanding extreme
performance, in terms of resistance per capacitance, junction arrays
resistors are a viable technology.\par 
\section{Superconductor-insulator transition in thin films} 
While the entry into the superconducting state below a critical
temperature $T_c$ is a thermodynamic phase transition, the
superconductor-insulator transition (S-IT) we will consider in this
section is 
(or `may be interpreted as')
an example of a quantum phase transition (QPT)
\cite{sondhi:97:qptrmp}. 
QPT take (in principle) place at zero temperature, and the S-I phase
boundary is crossed by varying a parameter other than the temperature in
the system's Hamiltonian. This can be the charging energy in a
Josephson-junction array \cite{geerligs:89:jjaprl,chen:92:ps} or the
amount of disorder in a metal undergoing a metal-insulator transition
(M-IT). Superconducting thin films have similarities with both
disordered metallic films and arrays of Josephson junctions
\cite{jaeger:89:scprb}. The small grain size generally means that even
charging effects are important in these films.\par
Experimental studies have been performed on quench condensed films, that
are films deposited from the vapour phase onto a very cold surface. This
technique allows to grow amorphous or nanocrystalline
films. Experimental studies on a variety of materials show that the
parameter governing the behaviour seems to be the sheet resistance
of the thin films. White et al. \cite{white:86:quenchprb} have measured
the superconducting gap with tunnelling experiments and found that the
broadening of the gap edges became comparable to the gap itself, and
hence superconductivity disappeared, when the sheet resistance at high
temperature reached (10\dots 20)\,k$\Omega/\Box$. \par
Jaeger \cite{jaeger:86:gallprb} found that gallium films became globally
superconducting when the sheet resistance was below about
6\,k$\Omega/\Box$. Experiments 
\cite{haviland:89:2dscprl,liu:93:sitprb}
suggest that the threshold for the
superconductor-insulator transition is a sheet resistance of one-fourth
the Klitzing resistance, the so-called quantum resistance for pairs
\begin{equation}
R_Q=\frac{\displaystyle h}{\displaystyle (2e)^2}=
\frac{\displaystyle R_K}{\displaystyle 4}\approx
6.45\,\mbox{k}\Omega.
\end{equation}
There is no conclusive 
agreement to date on whether this value is a
universal sheet resistance for the S-IT \cite{belitz:94:rmp}.\par
In all these experiments, the sheet resistance depended very sensitively
on the film thickness; in most cases, a difference of one nominal
monolayer can change the sheet resistance over the entire range of the
S-IT \cite{jaeger:89:scprb}. This problem was addressed by Wu
and Adams
\cite{wu:94:sitprl}, who used the same technique we used for the
experiments described in this report: the films 
(Al in his case, Nb in our experiments) were deposited with a
certain (relatively high) thickness and then thinned by controlled
anodic oxidation.
\section{Anodisation of niobium} 
Anodic oxidation or `anodisation' is the process of forming an oxide
layer by electrolysis on a metal anode in a suitable
electrolyte. Historically, the first major use of anodic oxide films was
as dielectrics in electrolytic capacitors
\cite{jackson:76:elcap,guentherschulze:37:elko}, 
and a lot of development work especially
with regard to electrolytes was carried out by industry and is hence
scarcely documented in the scientific literature. A general reference on
anodic oxide films is Young's 1961 book \cite{young:61:anox}, and
niobium is covered in detail in d'Alkaine's 1993 series of papers
\cite{alkaine:93:anodi1,alkaine:93:anodi2,alkaine:93:anodi3}. 
Halbritter \cite{halbritter:84:habil,halbritter:87:nbapa} treats the
subject of niobium and its oxides with focus on the interfacial
structure.
After
giving some information on the electrochemistry of niobium and its
oxides, we will take a look at 
various applications of anodisation for
micro- and nanofabrication of niobium that are documented in
literature. 
\subsection{Electrochemistry of niobium and its oxides}         
Niobium was discovered by Charles Hatchett in the year 1801
and originally named Columbium. In the
following years, it became confused with Tantalum, discovered 1802,
with which it occurs mostly in nature,
and was finally isolated and rediscovered
in 1844 by Rose and named
Niobium (after Niobe, the daughter of Tantalos). 
Both names were used until element 41 was officially named
Niobium by IUPAC in 1950, but the name Columbium is
to date still used 
occasionally by the American metallurgical community and e.\,g.
the United States Geological Survey. In metallic form, niobium was
isolated for the first time by v.~Bolton in 1905.\par
Early work on niobium anodisation was inspired by potential applications
in electrolytic capacitors
\cite{guentherschulze:37:elko,johansen:57:anox}, but Nb 
electrolytic capacitors
did not become a large scale commercial product like those based on
Ta. Today, niobium oxides are often studied because they form the surface
of superconducting accelerator cavities, and since acceleration is a
high frequency application, the surface is very important to the cavity
quality. Such cavities used to be made of sheet niobium, but are
nowadays also produced from copper covered with sputter deposited niobium
\cite{schucan:94:da}.\par  
There exist three stable oxides, niobium pentoxide Nb$_2$O$_5$, niobium
dioxide NbO$_2$, and niobium monoxide NbO, and  the solution of oxygen
\cite{seybolt:54:solu}
in niobium notated as Nb(O), with up to one weight percent of oxygen at
high temperatures. Niobium pentoxide occurs as NbO$_x$ with $x\in
[2.4\dots 2.5]$, the dioxide and pentoxide only in narrower
stoichiometry \cite{gray:75:iss}. Nb$_2$O$_5$ is the principal
constituent of anodic oxide films on niobium
\cite{bakish:60:anox,young:60:anox}. Its density in bulk amorphous form
is $\varrho=4360$\,kg/m$^3$, and its dielectric constant 
$\varepsilon\approx 41$ \cite{young:60:anox}; values for thin films
might deviate from this value, though.
Nb$_2$O$_5$ is an insulator, NbO a superconductor with 
$T_c\approx1.4$\,K.\par
The microstructure of an anodic oxide film on niobium is rather
complicated. The outermost layer is Nb$_2$O$_5$, followed by a thin
layer of NbO$_2$ followed in turn by NbO. This sequence was determined
by Gray et al. using ion scattering spectroscopy \cite{gray:75:iss}. It
is noteworthy that they found a more gradual falloff in stoichiometric
oxygen content from Nb$_2$O$_5$ to Nb on anodised foils than in natural
oxide layers. Halbritter points out that the interface between niobium
and its oxides is not even but serrated
\cite{halbritter:89:arxpseca}. This serration is stronger for `bad'
niobium as measured by the residual resistance ratio (RRR), the ratio of
resistivities at room temperature and at 4.2\,K
or just above the transition. Niobium deposited by
thermal evaporation is, compared to sputter deposited material, always
`bad', but evaporation in conjunction with a liftoff mask offers
more flexible patterning techniques.\par 
The reason for the serration of the interface is the volume expansion from
Nb to Nb$_2$O$_5$ by a factor of about 3 \cite{kroger:81:anodiapl} in
combination with the mechanical properties of the compounds involved. Nb
(density $\varrho=8570$\,kg/m$^3$ at room temperature)
is relatively soft, and niobium pentoxide microcrystallites cut into the
metal. This serration does not occur on the metals NbN and NbC that are
harder than Nb; carbon inclusion in the interface is known to improve
the quality of Nb based tunnel junctions \cite{kuan:82:nbox}.\par
\subsection{Micro- and nanofabrication by anodisation}
For purposes of micro- and nanofabrication, anodic oxidation is
basically used as a means to remove Nb metal; since the oxide is not
removable without affecting the underlying metal, it often becomes
an integral part of the design as an insulating layer.\par
\begin{figure}                                                  
\begin{center}                                                  %
\epsfig{file=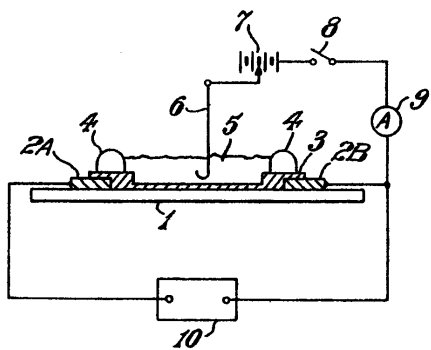,width=0.5\textwidth}          %
\end{center}                                                    %
\caption[Anodisation setup for resistor fabrication.]{
\label{anodisetupscanfig}Anodisation setup for resistor         %
fabrication, reproduced from Western                            %
Electric Company's 1959/1960 patent                                     %
specification \cite{westernelectric:60:resipat}.                %
(3) is a metal strip,                                           %
(10) a resistance monitoring device.}                           %
\end{figure}

Anodisation thinning of Nb metal strips (or other anodisable metals) for
the fabrication of tunable thin film resistors was patented to Western
Electric Company in 1959/1960 \cite{westernelectric:60:resipat}.
Figure~\ref{anodisetupscanfig} is a reproduction of a drawing from their
patent application, showing a setup quite similar to that used by us for
microanodisation described below (\ref{microanodisec},
cf. figure~\ref{microanodisetupfig}). 
A metal strip (3), deposited on a substrate (1), is covered with 
electrolyte (5)
confined by a mask structure (4). The resistance of the samples is
monitored via two leads (2) by some means (10), and the anodisation
voltage between the metal strip and the cathode (5) is adjustable (7,8),
e.\,g. such that the anodisation current (9) is kept constant. The latter
does not apply in our case; we will come back to the anodisation process
for resistor fabrication in detail in \ref{resimeascontrsec} and
\ref{anodidynsec}. \par
In large scale, anodisation is used for the production of small
junctions from prefabricated three layer sandwiches 
(Nb/AlO$_x$/Nb) through the
selective niobium anodisation process (SNAP
\cite{kroger:81:anodiapl}). The top layer metal is oxidised where a
photolithographically defined resist mask exposes it to the electrolyte,
and then suitable contacts are made to the bottom layer and the
unoxidised part of the top layer. This technique allows making high
quality junctions since the trilayer can be formed under conditions and
by methods (e.\,g. sputtering) that cannot be used for deposition
through a liftoff mask. Anodisation voltages are usually monitored to
determine the etch end.\par
Ohta et al. used anodisation to thin a weak link \cite{ohta:87:wljj} and
observed a transition to a tunnel Josephson junction in the temperature
dependence of the critical current, and Goto made variable thickness
bridges (VTB \cite{goto:79:jap,goto:81:vtb}) 
in Nb strips several micrometres wide by anodising through
a mask with an opening of only 200\,nm, produced from silicon monoxide
by shadow evaporation at a resist mask step. Here also a transistion
to Josephson junction behaviour was found, this time manifesting itself
in the occurence of Shapiro steps under microwave irradiation. Both Ohta
and Goto anodised without a monitoring device and assumed a certain Nb
consumption proportional to the applied voltage, an assumption whose
validity we shall inspect in \ref{anodidynsec}.\par
Anodisation fabrication and single electronics were combined
when Nakamura et al. made the so-called
anodisation controlled miniaturisation enhancement (ACME
\cite{nakamura:96:acme}) of single electron transistors in 1996. They
made single electron transistors in the `conventional' 
aluminium technique
with shadow evaporation and oxidation
\cite{niemeyer:74:mitt,dolan:77:masks} and
subsequently anodised the complete structure in order to minimise the
junction area, thus the capacitance, and raise the operating
temperature. In such a sample, they were able to observe a modulation of
the source-drain current with gate voltage up to temperatures of
30\,K. During the fabrication, they monitored the resistance through the
SET continuously, and were able to increase its value by two orders of
magnitude. Preexisting asymmetries of the junctions in the SET seem to
be enhanced by this process, so that most of the samples showed a
Coulomb staircase.
\chapter{Nanofabrication and microanodisation}  
\begin{figure} 
\begin{center}                                                  
\epsfig{file=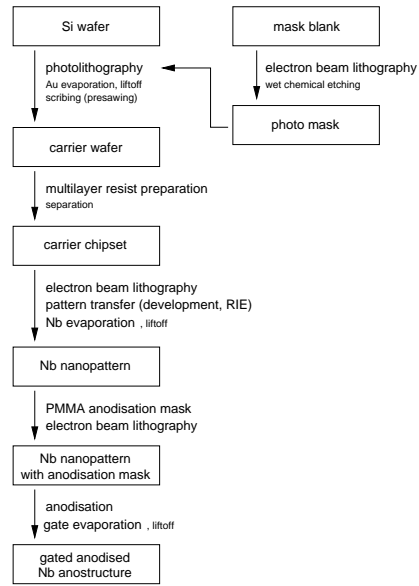,height=0.4\textheight}
\end{center}                                                    
\caption[Fabrication process: overview]{%
\label{fabovervfig}Fabrication of anodised Nb nanostructures    
with gates: process step flow scheme.}                  
\end{figure}

In this chapter, the techniques used for the nanofabrication and
microanodisation of niobium samples will be described. Figure
\ref{fabovervfig} gives an overview over the fabrication
process. Some of the information is of special interest with regard to
the equipment used, namely the JEOL JBX 5D-II electron beam lithography
system; most of the specific details, though, may be found in the recipe
appendix \ref{recipes}.
\section{Substrate and carrier system} 
Substrate material for all our samples are silicon wafers in (100)
orientation with a diameter of two inches and a thickness of about
0.25\,mm that have been thermally oxidised to an oxide thickness of
about 1\,$\mu$m. Since doping and resistivity of the silicon do not play
a role, wafers in sufficient quality can be obtained at a price of less
than 10\,ECU each. A disadvantage of oxidised silicon 
(compared to unoxidised silicon)
as a substrate
material for single electronics is that it makes components more prone
to damage from static electric discharges. On the other hand, it allows
testing of components at room temperature, saves time and chemicals
otherwise needed to remove the native oxide, and enables anodic
oxidation of materials deposited on the substrate without having to
worry about the anodisability of the silicon itself.\par
Standard chip size for our measurement equipment is $7\times 7$\,mm$^2$,
limited by the space  available in the commercial 
dilution
refrigerator
(see \ref{cryosubsec}). Cabling in this refrigerator limits the number of
contacts to sixteen, of which thirteen are dc leads and three coaxial
cables for rf signals. Contact between the leads and the chip is made by
spring-loaded probes (`pogo-pins').\par
The centre of each chip contains an area of $160\times 160\,\mu$m$^2$
where nanopatterns are defined by electron beam lithography. From this
area, dimensioned to equal four `fields' of the EBL system in highest
resolution mode (see glossary appendix \ref{glossary} for definitions),
gold leads connect to the contact pads for the pogo-pins situated about
two and a half millimetres away from the chip centre (a certain minimum
distance is advantageous for the subsequent anodisation step, to
facilitate placing of a droplet of electrolyte).\par
Besides the contact pads and leads, the gold chip pattern comprises a
chip number field, visible with the naked eye and used to orientate the
chip during handling, and four alignment marks (`wafer marks', see
glossary). This pattern is most economically created by
photolithography.\par 
A  process for carrier chip photolithography \cite[appendix to chapter
4]{chen:94:thesis} 
had to be abandoned to comply with  environmental
regulations restricting the use of toxic and carcinogeneous
chemicals. Instead, a new process was developed that not only eliminates
chlorobenzene but also requires less baking and chemical treatment
steps. A detailed account of this process is given in
\ref{padrecipe}. High quality in the photolithography is especially
important for the alignment marks, since several alignments are
performed during the whole fabrication process.\par
After photolithography, the wafers are presawed (`scribed') along the
chip edges from the back side and separated into sets of chips for
further handling; generally a set of two by two chips is processed at a
time. 
\section{A niobium liftoff nanofabrication process} 
\begin{figure}[t]\begin{center} 
\epsfig{file=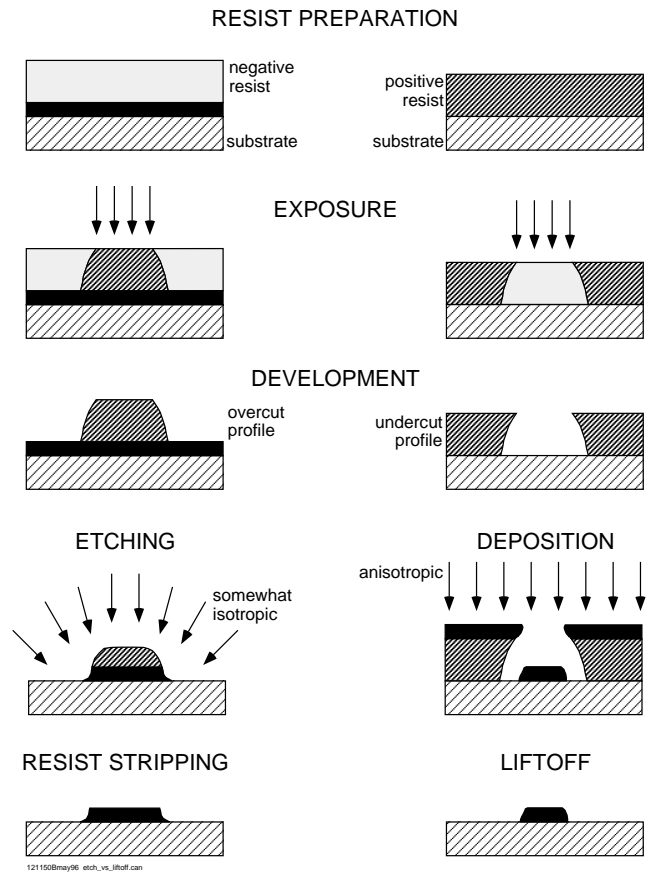,width=0.7\textwidth}
\end{center} 
\caption[Etch vs. liftoff process]{%
\label{etchliftofffig}Basic pattern transfer types relevant for
nanofabrication: etching and liftoff processes. Note the widening of the
exposure profile due to scattering of the electron beam, the resulting
overcut profile in negative and undercut profile in positive
resist. Isotropy of etching and mask erosion lead to a concave, clogging
of the liftoff mask on the other hand to a convex profile of the
patterned material in the respective processes.}
\end{figure} 
There are two basic pattern transfer process types relevant for
nanofabrication, namely etch processes and liftoff
processes. Figure~\ref{etchliftofffig} shows a comparison of these two.
In the case of an etch process, the material that is to be patterned is 
deposited on the whole substrate, covered with resist and etched away
where the resist has been removed after exposure and development.
In a liftoff process, the resist is patterned first, and the relevant
material deposited onto the patterned resist. It is removed
(`lifted off') together with the resist mask in a strong solvent.\par
A special form of liftoff process is the angular evaporation
technique \cite{niemeyer:74:mitt,dolan:77:masks}.\par
Making weak links in very thin niobium films by an etch process is
very difficult since native oxides cause a significant uncertainty in
the initial etch rate.
Since  the angular evaporation technique was assessed
essential for our samples in SET-like geometry
(and promises the possibility to make
overlap junctions), we decided to build know-how in a niobium liftoff
nanofabrication process suitable for the angular evaporation technique. 
\subsection{Pattern design and compilation}                     
Since the JEOL JBX 5DII system is a vector scan system with a scan step
of only 2.5\,nm in highest resolution mode, patterns can be designed as
vector drawings without regard to future pixellation. All patterns used
for this work were designed using the AutoCAD programme on a Digital VAX
workstation. They were exported as Drawing Exchange Format (DXF)
drawings and converted to JEOL01 code \cite{jeol01spezifikationen} by a
local programme and then via a sequence of conversion steps to JEOL's
scanner format used for the control of the EBL system.\par
Parallel to the drawing, information on pattern placement and exposure
doses is supplied in form of the so-called jobdeck and schedule files
(see glossary appendix \ref{glossary}).\par
By the time of this writing (April 1997), the JEOL01 format has been
largely obsoleted by the introduction of the PROXECCO proximity
correction programme \cite{eisenmann:93:proxecco} (see \ref{pdwebl}),
which produces output in the
industry standard Calma stream format (GDS II)\cite{gdsiientry}.\par
\subsection{Four layer resist system}                   
The melting point of niobium is 2468\,$^\circ$C, so that it can only be
thermally evaporated by electron gun heating and not from boats. Liftoff
masks are thus subjected to intense thermal load, and all-polymer masks
are generally regarded as unsatisfactory for Nb patterning. Even the
system of a germanium mask supported by PMMA or P(MMA-MAA) used
for the fabrication of ultrasmall tunnel junctions
\cite{geerligs:90:thesis}
is easily damaged.\par
\begin{figure}[t]                                                 
\begin{center}                                                  %
\epsfig{file=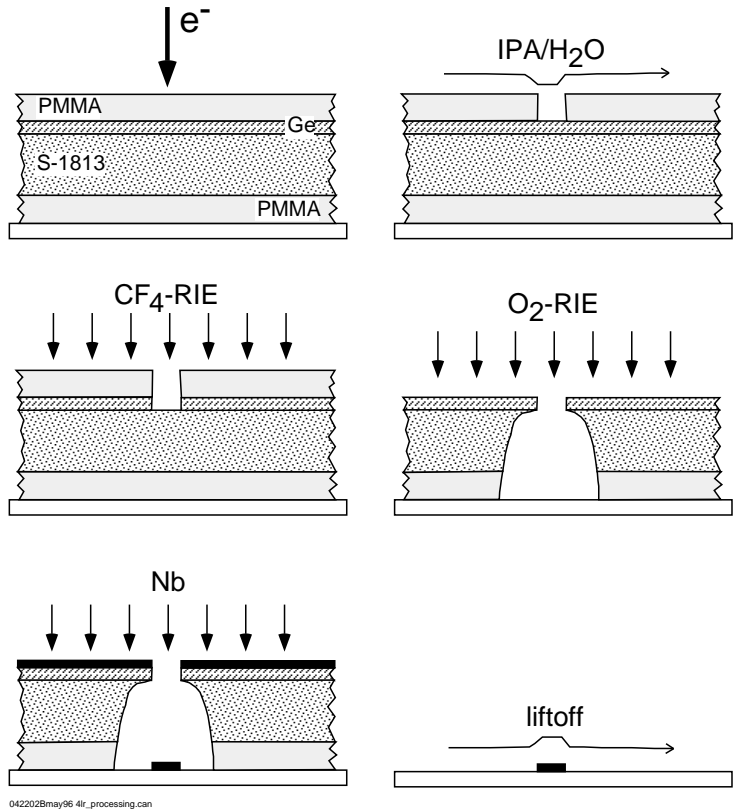,width=0.8\textwidth}      %
\end{center}                                                    %
\caption[Four layer resist processing]{\label{4lrprocessfig}
Four layer resist: overview of processing steps for Nb liftoff  %
nanopatterning. After exposure and development of the top       %
layer, the pattern is transferred to the germanium mask by      %
 rective ion etching with                                       %
carbon tetrafluoride. Subsequent RIE with                       %
oxygen creates the undercut                                     %
profile needed for a liftoff mask.}                             %
\end{figure}

A more suitable system is an aluminium mask on a polyimide support used
by Jain et al. \cite{jain:85:nbpbieee}. Liftoff is difficult, however,
since polyimide is quite resistant to strong solvents like
acetone. Harada addressed this problem by introducing the four layer
resist \cite{harada:94:nbset} whose structure can be seen in the top left
panel of figure~\ref{4lrprocessfig}: a germanium mask supported by a
layer of hard-baked photoresist S-1813 end equipped with a PMMA bottom
layer to enable liftoff. 
Like in many resist systems for
submicrometre lithography, the top layer patterned by EBL consists of
PMMA. 
\subsection{Pattern definition with electron beam lithography} 
\label{pdwebl}
Electron beam lithography is a pattern generation technique based on
physicochemical modification of matter under electron beam
irradiation. In its most usual form, an organic substance (`resist')
either increases or decreases in solubility in a proper agent
(`developer') under the influence of the e-beam. Resists are called
positive in the first and negative in the latter case. A certain dose is
required to achieve the modification, defining the sensitivity of the
resist. The ratio of the solution rates of exposed and unexposed resist
is called the contrast of the resist under certain process
conditions.
PMMA is a positive resist with a rather low sensitivity, but with good
contrast.\par 
An electron beam lithography machine generates a pattern by sweeping a
very fine beam over the sample, using magnetic lenses, and blanking the
beam where exposure is not desired. Much of the technology involved
stems from scanning electron microscopy, and the cheapest EBL machines
are indeed SEMs with some extensions. Our JEOL system, however, is a
commercial machine designed for high precision lithography 
on large areas, which 
was very useful for the quantum dot work described in the
appendix omitted from this WWW
edition. All patterns were written in the highest
resolution mode (for insiders: first aperture, fifth lens). The lowest
current, giving the smallest beam diameter, was 20\,pA, and the maximum
current 
used, one that could be achieved reliably even with an aged emitter, was
1\,nA. These two currents were used for lithography of the fine patterns
and coarse leads, respectively. Division of patterns into low and high
current patterns requires some additional adjustment time, but can give
substantial savings in exposure times, thus allowing for numerous test
patterns. Such test patterns with varied dose are valuable tools for the
assessment of development process steps.\par
Finding the right dose for a pattern is a nontrivial problem due to the
proximity effect. Forward scattered, backscattered, and secondary
electrons place an electron dose outside the area directly hit by the
electron beam. For large beam diameters, even the nonuniform current
density in the beam (assumed to be a Gaussian) must be taken into
account. As a consequence of the proximity effect, large areas
consisting of many pixels require a lower 
averaged irradiation dose because
surrounding pixels contibute to the dose in a certain spot. 
Conversely, small patterns need to be exposed with a higher dose in
order to develop properly. There are in principle three ways of
performing the proximity correction, i.\,e. the assignment of different
doses to pattern details of different width and in different
surroundings:
\begin{enumerate}
\item Trial-and-error: based on some initial guess, test exposures are
performed and the results assessed by electron microscope inspection.
\item Rules-of-thumb: proximity correction can be estimated using
e.\,g. the formulas given in \cite[appendix C]{delsing:90:thesis}.
\item Correction programmes: a number of commercial software products
for proximity correction are available that calculate local correction
factors from CAD data and a suitable correction function. Such a
correction function, in turn, is generated by simulation programmes,
usually using a Monte Carlo approach.
\end{enumerate}
Most of the patterns described here were designed using manual proximity
correction with the trial-and-error method, leading to the doses given
in \ref{4lrexposurerecipe}.
Recently (March 1997),
however, the transition to PROXECCO-corrected patterns 
has been
initiated. Simulations are performed using a local programme aptly named
\texttt{mcarlo}.
Relevant parameters are given in \ref{proxparapp}.\par
\subsection{Pattern transfer}
A simple developer for PMMA is a mixture of isopropanole and water. As
soon as possible after exposure, the patterns were developed using the
concentration, times etc. given in \ref{flpprcapp}. 
\subsubsection{Pattern transfer to the metal mask}
The openings in the EBL-patterned PMMA top layer were transferred to the
germanium layer by reactive ion etching (RIE). RIE is a combination of
chemical etching and physical sputtering. The sample is placed inside a
low pressure reaction chamber on an insulated electrode, and a reactive gas
(`process gas') is let into the reaction chamber.
A radio
frequency cold plasma discharge is then ignited in the chamber.
The process gas becomess
partially cracked by the discharge, creating highly reactive radicals
that can reach the sample because their mean free path is long enough in
the low pressure. These radicals provide the chemical etching, which is
essentially isotropic. Additionally, molecules accelerated in
the electric field have a sputtering effect. The lower the chamber
pressure, the higher is the anisotropy of this sputtering. Near the
electrode on which the sample is placed, a dc bias voltage occurs
between the electrode and the plasma potential due to different
mobilities of positive and negative ions. The anisotropy of the
etching increases with this dc bias.\par
A suitable reactive gas for germanium etching is carbon tetrafluoride
CF$_4$. Our RIE system was not equipped with an etch end detection, so
that the required etching time had to be estimated 
based on experience,
allowing some extra margin since reactive ion etching processes tend to
be somewhat irreproducible. Etch rates may vary depending on
contaminations present in the chamber,
or on the size of the sample areas, just to name a
few factors. Overdoing this etch step resulted in a slight increase in
linewidths, but this increase was considered tolerable.\par
Successful etching of the germanium layer is clearly visible by optical
microscope inspection. While developed PMMA areas appear just slightly
brighter than their surroundings in an optical microscope, the etched Ge
areas are much darker.
\subsubsection{Etching of the support layers}
Both the hardbaked photoresist and 
the PMMA bottom layers were etched with RIE
using oxygen as reactive gas. A higher pressure than in the Ge etch step
was used to increase the anisotropy of the etch rate to create the
desired undercut profile. Another advantage of higher pressure is that
the physical sputtering, leading to damage of the suspended Ge mask
parts, is reduced.\par
\begin{figure}                                                  
\begin{center}                                                  %
\epsfig{file=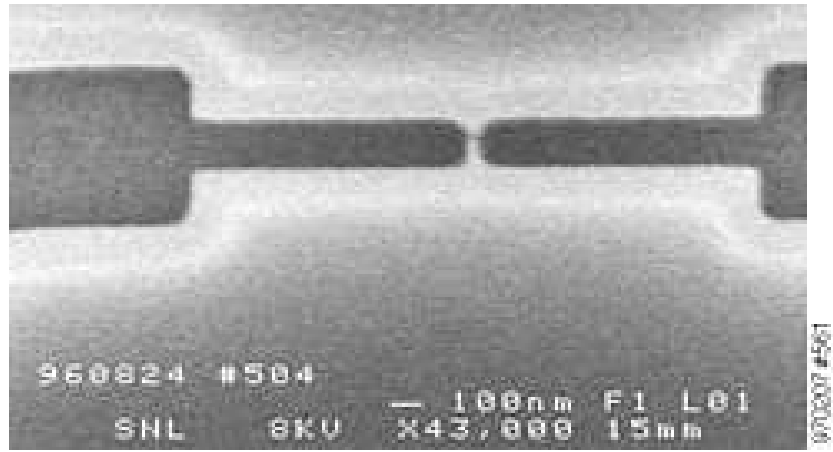,width=0.9\textwidth}         %
\psfull
\end{center}                                                    %
\caption[Four layer resist mask after oxygen etching]{
\label{fourlressem}Four layer resist mask after pattern         %
transfer and reactive ion etching steps. 
The darkest regions are the %
substrate (oxidised silicon), adjecent light areas      %
show the undercut created in the oxygen etching step. The       %
bridge in the center is damaged by a tiny crack.}               %
\end{figure}

Other than in the Ge etch step, the lack of an etch end detection is
rather
unfavourable here, because unnecessarily
long etching causes damage
of the Ge mask
that could otherwise have been avoided. 
Figure~\ref{fourlressem} is a scanning electron
micrograph of an etched four layer resist mask. The undercut is clearly
visible if one uses acceleration voltages of (5\dots 8)\,kV. The
suspended bridge in this picture is damaged in the form of a tiny
crack. More careful timing of the oxygen etch step and use of Teflon
piedestals to adjust the position of the sample in the reaction chamber
can reduce the risk of such damage.\par
\subsubsection{Surface quality and contamination}
\begin{figure}                                                  
\begin{center}                                                  %
\epsfig{file=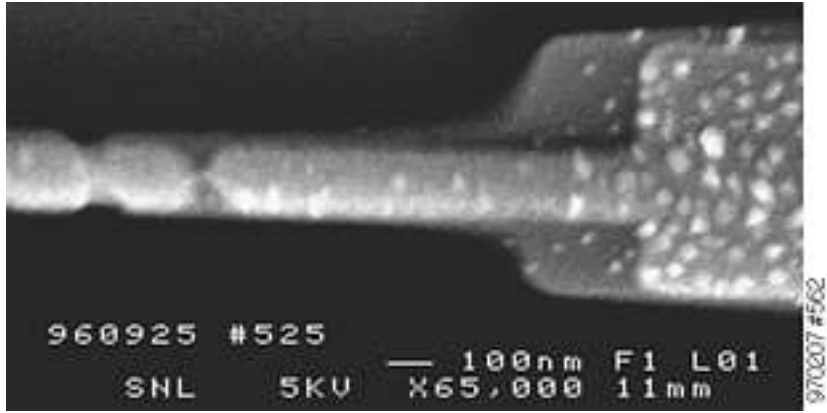,width=0.9\textwidth}         %
\psfull
\end{center}                                                    %
\caption[Contamination of large Nb areas]{
\label{contaminapic}Grainy contamination occasionally           %
observed on large Nb areas (right).}                            %
\end{figure}

The major problem associated with pattern transfer entirely by plasma
etching is that of surface contamination. Figure~\ref{contaminapic}
shows an example of grainy surface contamination that is frequently
observed on our samples, and is reported (at least 
inofficially) by other
users of RIE-only pattern transfer. 
The exact origin of this contamination is unknown. Gentler etching seems
to reduce this contamination, and corroborates the assumption that it is
caused by the redeposition of sputtered material during the oxygen etch
step. Since the contamination occurs mostly in wide open areas and
hardly affected line shaped structures below 200\,nm width, it could be
tolerated for our purposes.\par
Recently (March 1997), a mask based on a polyimide bottom layer was
introduced at the Swedish Nanometre Laboratory. The final pattern
transfer step here is a wet chemical development, and this process might
be an alternative to the four layer process described above.\par
\subsubsection{Angular evaporation of niobium}
The evaporation of good niobium requires large power and ultra high
vacuum (UHV). The first requirement is due to the high melting point and can
be met with the use of electron gun heating. The latter requirement,
however, is harder to meet since under the immense heat during
evaporation, most UHV systems suffer from outgasing. For the samples
described in this report, a 
non-bakable multipurpose HV system was used, 
not  a UHV system, and reasonably low
pressure had to be achieved by a series of preevaporations and long-time
pumping. Still, the pressure during evaporation was typically about
$(3\pm 1)\cdot10^{-5}\,$Pa, which gives pretty poor niobium
with a $T_c$ of the order of 1.5\,K. A new
evaporation system tailored to the specific needs of niobium evaporation
is presently under test
and gave films with a $T_c$ around 9\,K in 100\,nm thick films.\par
The film thickness was monitored during evaporation with a conventional,
water-cooled crystal resonance bridge, and controlled ex situ with an
Alphastep stylus-method profilometer. For a typical film thickness of
20\,nm, the accuracy of the thickness control appears to be not better
than 10\,\%, which is a rather strong limitation for the fabrication of
double weak link structures depending on the symmetry of the two weak links.
\begin{figure}\begin{center} 
\epsfig{file=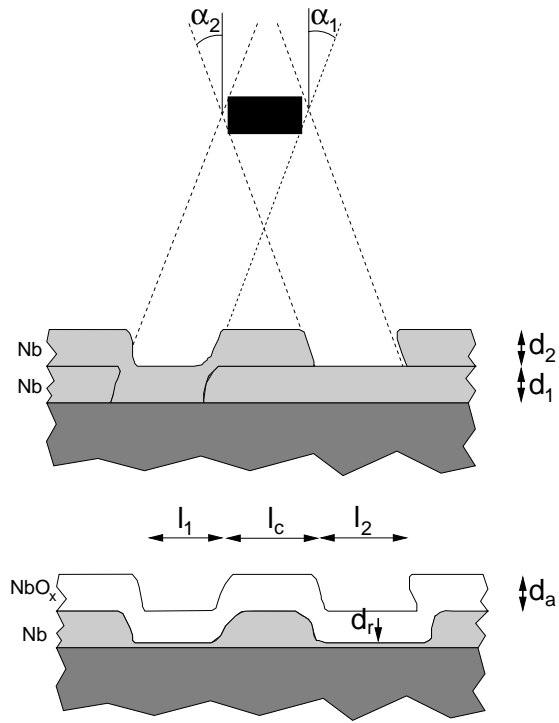,width=0.6\textwidth}
\end{center}\caption[Technique for fabrication of
samples in SET-like geometry]{\label{shadowsketch}%
Fabrication of samples with two weak links in single electron
transistor-like geometry. Angular evaporation with a suspended bridge
mask (top) defines the two weak links separated by an island. After
anodisation (bottom), the remanining Nb is much thinner in the weak
spots. Neglected in this sketch are the deposition of material on the
mask and the swelling of the film during anodisation.}
\end{figure} 
Figure~\ref{shadowsketch} is a schematic drawing of the angular
evaporation process we used for the fabrication of samples with two weak
links in SET-like geometry. In this approach, the length $l_1\approx
l_2$ of the weak links is defined by lithography (as the width of a
suspended bridge), while the island length $l_c$ is created by an
overlap and can in principle be made very small. Subsequent anodisation
(bottom panel) creates an oxide layer with thickness $d_a$, thinning the
weak links to an averaged thickness $d_r$. In this sketch, we have
neglected the deposition of material on the suspended bridge, leading to
an asymmetry in the shape of both weak links, and the swelling of the
film during anodisation.\par
\begin{figure}                                                  
\begin{center}                                                  %
\epsfig{file=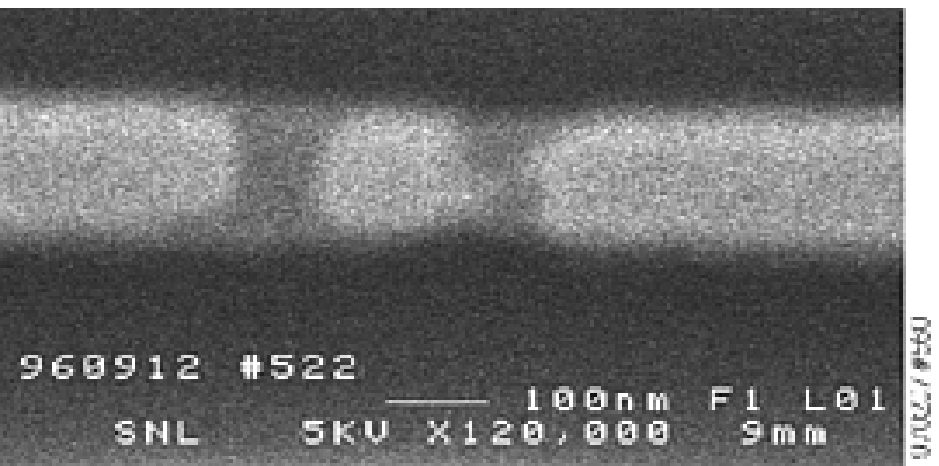,width=0.9\textwidth}         %
\psfull
\end{center}                                                    %
\caption[Variable thickness weak links in Nb wire]%
{\label{wealisem}Variable               %
thickness weak links in a Nb wire made by double angle          %
evaporation technique, after deposition and before further      %
processing.}                                                    %
\end{figure}                                                    %
A scanning electron micrograph of such a double weak link structure,
consisting of two 20\,nm thin spots in an otherwise 40\,nm thick film,
is given in figure~\ref{wealisem}. This sample was made by evaporating
the niobium under angles of $\pm 22^\circ$ to the substrate normal.\par
\section{Microanodisation of niobium}                           
\label{microanodisec}
In the following section, we will first introduce the equipment and
processes used for anodisation, and then present results on anodisation
dynamics that are of direct importance to our goal of fabricating tuned
resistors and weak link structures. We shall especially take a closer
look at oxide growth under constant voltage bias.
\subsection{Experimental setup and procedures}                  
\begin{figure}                                                  
\begin{center}                                                  %
\epsfig{file=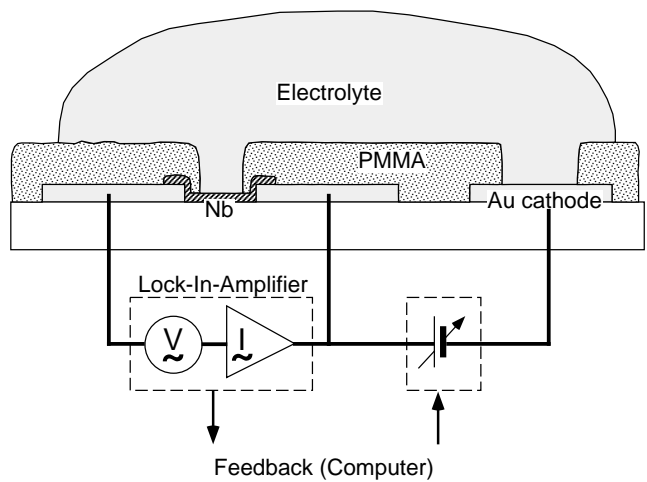,width=0.7\textwidth,
bbllx=34,bblly=108,bburx=542,bbury=497,clip=}                   %
\end{center}                                                    %
\caption[Experimental setup for Nb microanodisation]{
\label{microanodisetupfig}Experimental                          %
setup for microanodisation                                      %
of niobium nanostrips (schematic). In reality, electrical contacts are
placed at the chip perimeter, well away from the electrolyte droplet.}  
\end{figure}                                                    %
Figure~\ref{microanodisetupfig} shows a schematic drawing of our setup
for niobium microanodisation. In comparison with the historic drawing
(see figure~\ref{anodisetupscanfig}), the important differences are a
microfabricated anodisation mask, a cathode integrated on the chip, and
a specified means of resistance monitoring.
\subsubsection{Anodisation mask design and preparation}         %
The anodisation mask consists of a simple PMMA layer, at least
1.5\,$\mu$m thick. It meets the requirements for stability against
dielectric breakthrough under the required cell voltages 
(of up to almost
30\,V), can 
relatively easily be applied and patterned, and removed. It can even be
used as liftoff mask for gate electrodes, because a slight undercut
profile is created in such thick resists by backscattered electrons.\par
\begin{figure}                                                  
\begin{center}                                                  %
\epsfig{file=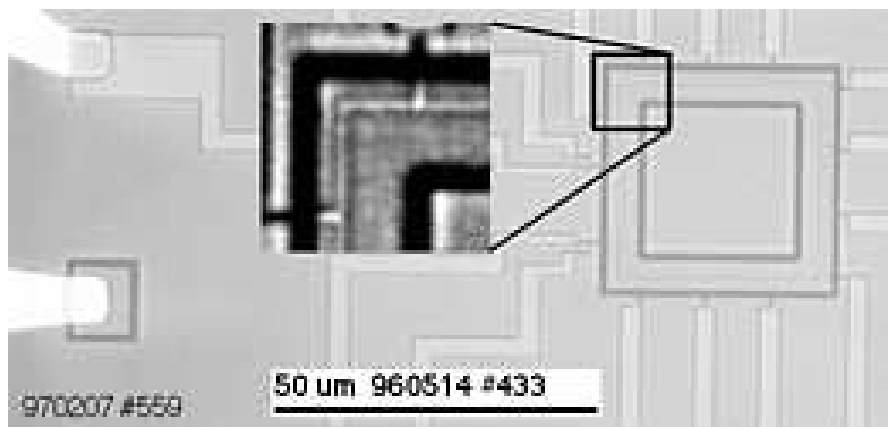,width=0.96\textwidth}        %
\psfull
\end{center}                                                    %
\caption[Anodisation mask, optical micrograph]{
\label{maskphoto}Anodisation mask, optical photograph.          %
Light structures on the left side are gold contact fingers,     %
the lower one                                                   %
being used as on-chip-cathode. The square frame                 %
shaped window on the right exposes the Nb wire, visible as      %
a thin white line with two leads
in the detail magnification.}                   %
\end{figure}                                                    %
An optical micrograph of an anodisation mask, here over a long resistor
wire sample, is given in figure~\ref{maskphoto}. The photo shows the
window over the wire, to the right, and another window over a gold
contact lead on the left. This gold lead served as cathode. Since the
amount of chemicals involved is  small 
even in relation to the size of the
electrolyte drop, chemical processes at the cathode have no major
importance, and the cathode can be made of practically any metal that
does not form an insulating surface oxide; niobium does not work 
well. The integrated cathode very close to the chip centre means that
only a tiny drop of electrolyte is needed, in practice the smallest drop
one can create with a syringe and place by hand. The need of inserting a
cathode into the electrolyte is eliminated, and the chip can be moved in
the sample holder with the electrolyte in place, which is a great
advantage. 
\subsubsection{Electrolyte}
As mentioned before, 
a lot of the development of suitable  electrolytes has
been done in industry many decades ago and is poorly documented in the
scientific literature. Many recipes are based on experience rather than
a deep understanding of the process details. Demands for a good
electrolyte are stability and a low vapour pressure. In the case of an
electrolytic capacitor, this reduces the need for hermetically sealed
encapsulation, and it is a very welcome property for us, since a change
in composition during 
the anodisation due to
exposure to atmoshere is undesirable.\par
A very important property of an electrolyte for anodic oxidation is that
it should be incorporated as little as possible into the oxide
film. More precisely, instead of anodic oxide film (AOF) one speaks of
anodic film (AF) \cite{delloca:71:aof}, when the incorporations from
the electrolyte are taken into account.\par
Niobium has the rather pleasant property that it forms a nonporous oxide
with many different electrolytes \cite{guentherschulze:37:elko}, unlike
valve metals 
with incomplete valve action
(``unvollst\"andige Ventilwirkung'', \cite{guentherschulze:37:elko})
like aluminium. In the latter case, many electrolytes have
a solving effect on the formed oxide, leading to a porous oxide
structure. Contamination, especially by halides, is also known to
increase the porosity of aluminium AF \cite{delloca:71:aof}.
For some applications, this is exactly desired, like for the
surface treatment of aluminium where pigments are brought into the
pores, but for electrolytic capacitors, a strong uniform 
and thin oxide is
essential.\par
Possible electrolytes for niobium anodisation are saturated boric acid
\cite{chiou:71:timedep} or an aqueous solution of sodium tetraborate and
boric acid \cite{aponte:87:barr}. We used, however, an aqueous solution
of ammonium pentaborate mixed with ethylene glycole. These compounds
have been in use since the 1940's, and we could trace back our exact
recipe to a 1967 paper by Joynson \cite{joynson:67:anodi}. Originally
used for the forming of pinhole-free oxide films, the same electrolyte
was used by Kroger for the selective niobium anodisation process
\cite{kroger:81:anodiapl}. The electrolyte was designed for use at an
elevated temperature of 120$^\circ$C \cite{joynson:67:anodi}, but works
in principle at room temperature, if it is stirred and heated to about
100$^\circ$C for a few minutes, not more than two days prior to
use. After two days, precipitation occurs. For niobium anodisation, the
prepared solution could be used for at least one year without any
noticeable change in properties. The stability and operability over a
large temperature range are typical properties of a good capacitor
electrolyte. \par
Ethylene glycole seems to have several positive effects: it influences
oxygen chemisorption and promotes the formation of more stoichiometric
oxides \cite{bairachnyi:92:gly}, and this particular electrolyte is
known to show little incorporation of electrolyte matter into the anodic
film \cite{santway:70:gly}. Side reactions occur at higher voltages
\cite{santway:70:gly} and appear negligible for our work, where 30\,V
is never exceeded.\par
\subsubsection{Resistance measurement and control}              
\label{resimeascontrsec}
The requirements for a resistance monitoring device are:
a certain
precision, the ability to work in spite of the potential difference
between the sample and the electrolyte and cathode, and a negligibly
small distortion of the anodisation field to avoid systematic skewing of
the anodisation profile. All these requirements are met by using a
lock-in amplifier.\par
We used a Stanford SR850DSP amplifier in current measuring mode,
applying a sine shaped excitation with typical rms amplitudes of 4\,mV,
for very high resistive samples up to 32\,mV. With a suitably chosen
time constant, resistance measurement accuracies of a few percent can be
achieved. Since the cabling of the sample holder was quite open and
pickup non-negligible at low frequencies, a measurement frequency of
3000\,Hz was used for most of the samples mentioned here.\par
Careful grounding of the measurement equipment and of the syringe when
applying the drop to the chip are strongly recommended.\par
\begin{figure}                                                  
\begin{center}                                                  %
\epsfig{file=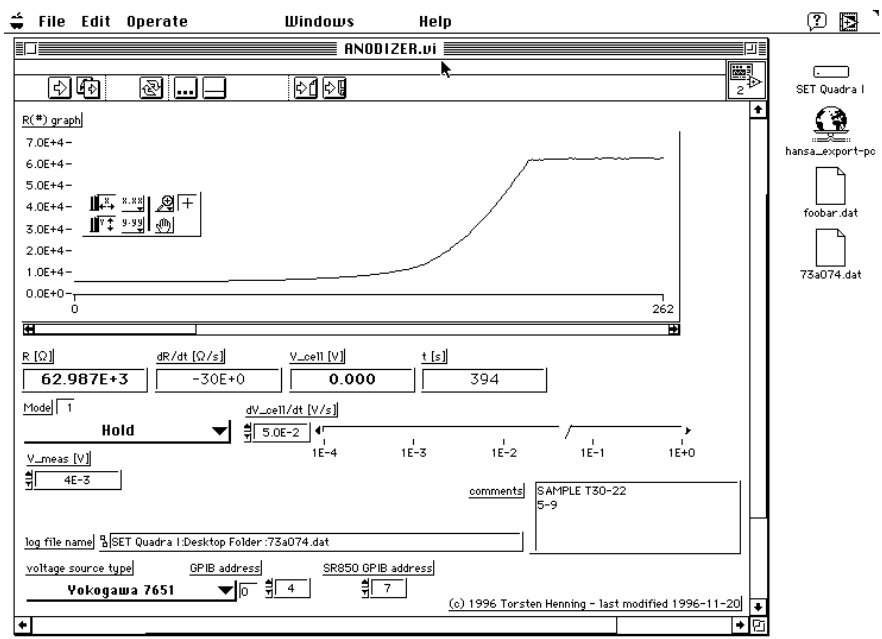,width=0.96\textwidth}
\psfull
\end{center}                                                    %
\caption[Anodisation control programme user interface]{%
\label{anodizerwd}Graphical user interface of the resistance monitoring
and anodisation voltage control programme. The main display shows the
resistance as a function of time.}
\end{figure}
\begin{figure}[t]                                                       
\begin{center}                                                  %
\epsfig{file=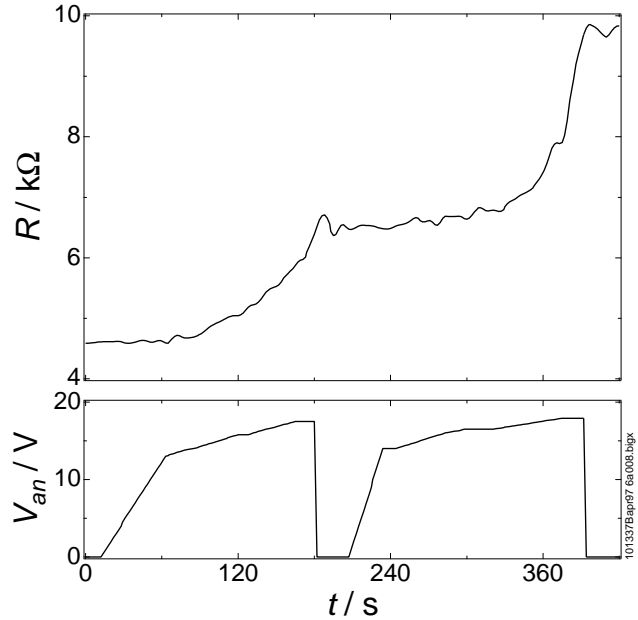,width=0.7\textwidth}      %
\end{center}                                                    %
\caption[Anodisation process]{\label{formfig}Anodisation process.       %
The resistance $R$ along the wire sample increases irreversibly %
with the anodisation voltage $V_{\it an}$ and with time.}       %
\end{figure}                                                    %
Resistance and anodisation value readings were taken automatically via a
GPIB equipped Macintosh  and logged. A graphical user
interface implemented in LabVIEW allowed the operator to monitor the
resistance value and adjust the anodisation voltage ramp rate
accordingly. Figure~\ref{anodizerwd} is a screen dump of this user
interface, showing the resistance monitor display and controls for ramp
mode and ramp rate and demonstrating the degree of control and
smoothness of the resistance tuning.\par
The sample resistance increases irreversibly with time as the anodisation
voltage is applied. The rate of resistance increase in turn is a
complicated function of voltage and time itself. In
figure~\ref{formfig}, the time evolution of 
the resistance and 
the anodisation voltage for one specific sample
are plotted. The resistance values are low since this was a weak link
sample, and most of the resistance was contributed by contact resistances
in the two probe measurement configuration. In spite of the somewhat
noisy data, one can see that the resistance along the sample (top panel)
remains constant as the anodisation voltage (bottom) is zeroed.\par
\subsection{Anodisation dynamics}                               %
\label{anodidynsec}                                             
\begin{figure}                                                 
\begin{center}                                                 %
\epsfig{file=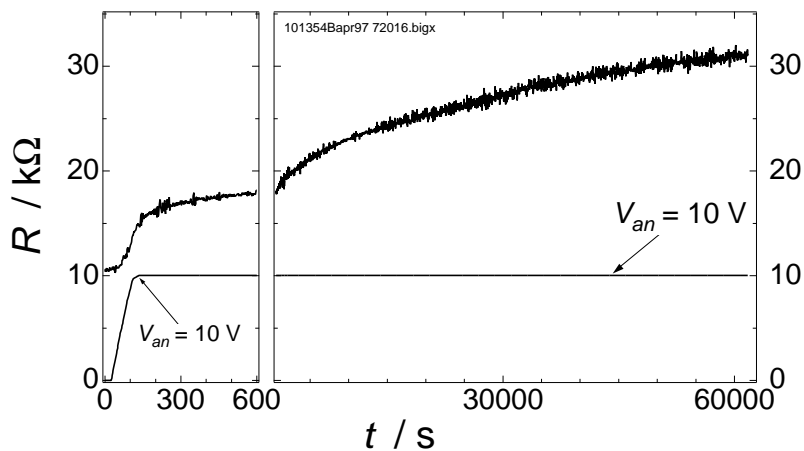,width=0.9\textwidth}     %
\end{center}                                                   %
\caption[Long-time anodisation process]{\label{longformfig}
Long time anodisation process. As the                              %
anodisation voltage is ramped to 10\,V                         %
(lower trace), the sample resistance rises                     %
rapidly in the beginning (left panel). The anodisation         %
voltage is then held constant,                                 %
and the sample resistance continues to increase                %
over more than sixteen hours,                                  %
demonstrating that there is                                    %
no limiting thickness for anodic oxide grown at                %
constant voltage.}                                             %
\end{figure}                                                   %
As mentioned before, the notion of an `anodisation constant', i.\,e. a
linear dependence between anodisation voltage and the thickness of the
metal consumed, does not hold in cases where the current density is not
kept constant all the time. Since our anodised areas were so small that
the anodisation currents were not accessible to measurement, we 
only measured the resistance along the sample and fixed
(or ramped) the anodisation voltage instead. The nonlinear dependence of
oxide growth on voltage and time is well illustrated by
figure~\ref{longformfig}. In the process depicted there, the anodisation
voltage was ramped up to a fixed value (10\,V) in 100\,s and then kept
constant for more than sixteen hours. All the time, the resistance
continued to increase measurably. The rate decreased, but under these
conditions, the oxide 
would theoretically continue to grow 
until the strip were completely oxidised. 
This effect can be exploited for the precision tuning of
resistors. By stopping the voltage ramping well below the desired resistance
value and simply waiting and zeroing the voltage as the desired
resistance is reached, one can achieve a resistance tuning whose
accuracy is in principle only limited by the measurement accuracy of the
resistance monitor.\par

\section{Top gate electrodes}
The deposition of the overlapping gate electrodes onto the samples in
SET-like geometry was the last fabrication step. After some notes on the
fabrication itself, we will take a look at the insulation properties of
the anodic oxide.
\subsection{Preparation}
\begin{figure}                                                  
\begin{center}                                                  %
\epsfig{file=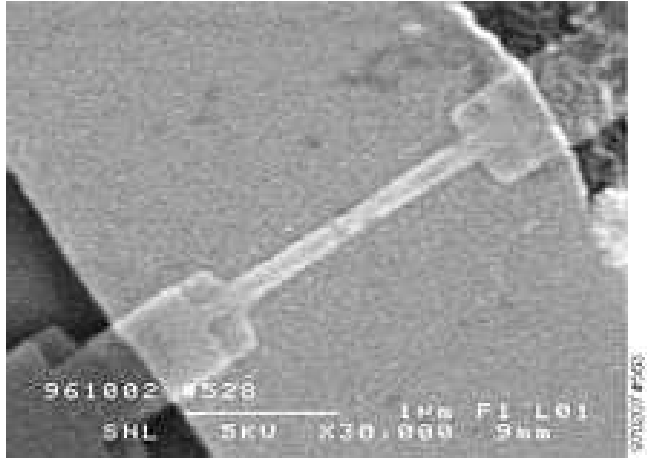,width=0.7\textwidth}         %
\psfull
\end{center}                                                    %
\caption[Top gate on anodised double weak link structure]{
\label{topgatesem}Gold top gate on an anodised sample           %
with two weak links (barely visible in the                      %
centre), following the contours of the                          %
anodisation mask.}                                              %
\end{figure}                                                    %
Making the overlapping gates was quite straightforward. The samples were
carefully rinsed with deionised water as soon as possible after the
anodisation process. The use of ultrasonic excitation seems to have a
negative effect on the anodised samples, most of them were destroyed
when such a cleaning process was tried out. A gentle surface ashing with
oxygen RIE was found to provide sufficient adhesion of gold to the
samples. 50\,nm Au were evaporated to ensure continuity over the sample
edges. These gates had a two terminal resistance of about 60\,$\Omega$
at a length of 160\,$\mu$m and an initial width of 8\,$\mu$m, narrowing
down to to 3\,$\mu$m over the samples. Figure~\ref{topgatesem} is a
scanning electron micrograph of a top gate deposited on a SET-like
sample. Contaminations were no major problem, in spite of the fact that
the present implementation of the anodisation technique in our
laboratory involves no special efforts to eliminate particle contamination.
\subsection{Electrical properties}
The most obvious demand in using a combined anodisation and gate liftoff
mask is of course that no metal be deposited on 
non- (or not sufficiently)
anodised areas. This requires a good wetting of the niobium by the
electrolyte, and indeed we found no electrical shorts or
failures to anodise niobium that had to be attributed to insufficient
wetting. The narrowest mask structures were about 3\,$\mu$m wide, and
every anodisation window was at least 8\,$\mu$m wide in one
direction.\par
In all but very few samples that we tested at low temperatures, the
insulation between the gate electrode and the sample was found to be
very good.
\begin{figure}\begin{center}
\epsfig{file=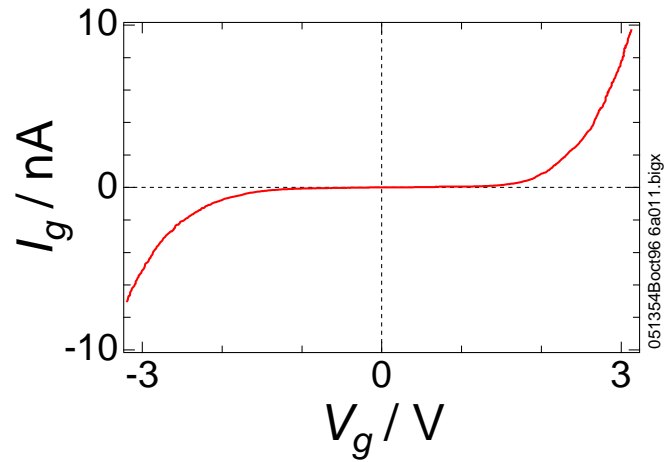,width=0.8\textwidth}
\end{center}
\caption[Top gate electrical leakage]{%
\label{topgateleakfig}Current-voltage characteristic for the leakage
current between gate and source-drain of a sample in SET-like geometry.}
\end{figure}
Figure~\ref{topgateleakfig} gives an example of a current-voltage
characteristic for the leakage current through the gate. Up to gate
voltages of about 1.5\,V, the gate current was too small to be measured
with our equipment, which translates into an insulation resistance of
better than 30\,G$\Omega$. 
For higher gate voltages, a measurable current flow set in. 
For other samples, this point lay at about
0.5\,V. In
transport measurements through the sample's drain and source, this
resulted in a shift of the measured IVC and was easily detected.
\chapter{Electronic transport in anodised niobium
nanostructures} 
After introducing the measurement setup common to all experiments, we
will present the results of transport measurements on resistor wire
samples and on samples with two weak links in single electron
transistor-like geometry.
\section{Measurement setup and procedure} 
As mentioned in the introduction, measurements on single electron
devices may
require very low temperatures,
and sometimes shielding from the (high
frequency) electromagnetic environment. We will briefly take a look at
our measurement setup under these aspects.
\subsection{Cryogenics} 
\label{cryosubsec}
The first measurements were performed in a noncommercial dilution
refrigerator that reached temperatures down to 95\,mK. The temperature
could be measured with a  resistance thermometer calibrated over the
whole accessible temperature range. Most measurements documented in this
report were, however, done in a commercial dilution refrigerator of type
TLE\,200 from Oxford Instruments. A 
germanium resistance thermometer was
calibrated down to 45\,mK, and the base temperature of the cryostat was
below 20\,mK, as estimated from a preliminary measurement with nuclear
orientation thermometry. A magnetic field up to 5\,Tesla could be
applied perpendicular to the sample. With a ramp rate of
0.1\,T/min, the sample was warmed to approximately 40\,mK by eddy
currents.
\subsection{Shielding and Filtering} 
The cryostat was placed in a steel enclosure
(`shielded room'). Inside this enclosure, all
electronics was analogue. Leads into the shielded
room were passed through filters in its wall. Inside the cryostat, the
leads were multiply filtered. The most prominent part of the filter
design are Thermocoax coaxial cables whose excellent filter properties
were pointed out by Zorin 
et al. \cite{zorin:95:thermocoax}. More information
about the cryostat and the filter design and properties can be found in
\cite{haviland:96:jvst}. 
\subsection{Measurement electronics} 
From the sample, the DC leads went via said Thermocoax
filters and multiply thermally anchored wires to an amplifier box on
top of the cryostat at room temperature. The bias voltage was
symmetrised with respect to ground and fed to the samples via high ohmic
resistors in the mentioned amplifier box. Voltage drops over the sample
and over the bias resistors (proportional to the current) were picked up
by low noise amplifiers, and the amplified voltages sent outside the
shielded room for registration. Details about the measurement
electronics can be found in  Delsing's PhD thesis
\cite{delsing:90:thesis}. \par
Signals were measured with digital voltmeters,
initially with DMM of type Tektronix  DM5520  with a buffer capacity of 500
points, later with Keithley 2000 DMM storing 1024 data points. The
measurement times were synchronised with a Keithley 213 voltage source
providing the bias voltage, which was stepped rather than swept
continuously. Gate voltages were either generated with a second port on
this Keithley 213 source, or with a Stanford 
Research Systems DS\,345 signal
generator.\par
Unless explicitly mentioned otherwise, the sweep of the bias or the gate
voltage was always bidirectional, starting and ending at one of the
edges of the swept voltage region.\par
All data were registered electronically with the help of a GPIB equipped
Macintosh. For historical and practical reasons, the
measurement software was written in various versions of
LabVIEW. Therefore, it cannot be documented in print. Binary data
files were transcripted into ASCII, and evaluated mainly using the
application software Igor from Wavemetrics.\par
\section{Resistor samples} 
After introducing the geometry of the resistor samples, we will go
through the different kinds of current-voltage characteristics, 
and present
our evaluation method for the quantitative analysis of the observed
Coulomb blockade. Finally, we will arrive at a criterion for the onset
of the Coulomb blockade. Via the Coulomb blockade, we observed a
superconductor-insulator transition.
\subsection{Sample geometry} 
The standard geometry for 
the resistor samples were strips of 10\,$\mu$m length
and a width limited by the lithography and 
pattern transfer technique in the respective
state of the art, that was between 120\,nm and 180\,nm
at the time of the resistor sample fabrication. These wires were
either single, and attached to wider Nb contacts for four probe
measurements, or grouped into a 120\,$\mu$m long wire as in
figure~\ref{maskphoto}. These wires had probe leads spaced at 10\,$\mu$m
distance from each other, and resistances were also measured in four
probe configuration.\par
\subsection{Current-voltage characteristics} 
\begin{figure}[t]\begin{center} 
\epsfig{file=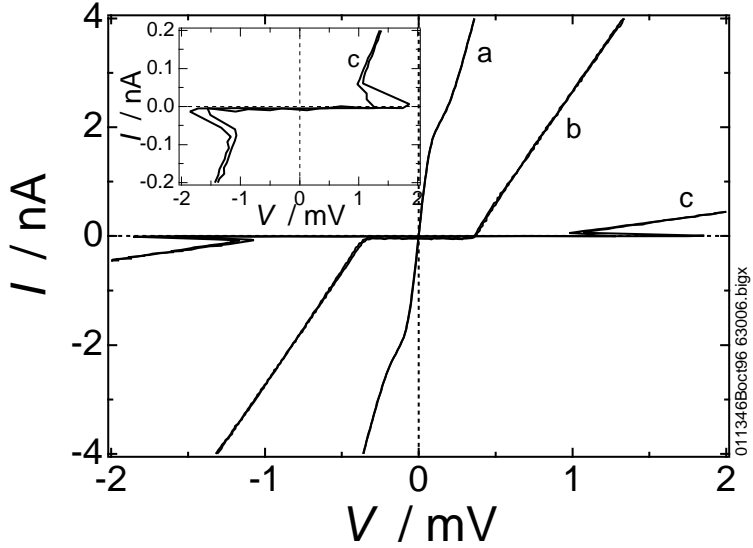,width=0.9\textwidth}      
\end{center}                                                    
\caption[Current-voltage characteristics for resistor samples]{%
\label{resiivcfig}Resistor sample current voltage               
characteristics (IVC). For different samples with the same      
dimensions ($10\times 0.15\,\mu$m$^2$), the IVC changes         
with increasing sheet resistance from a supercurrent            
remnant (a: 1.5\,k$\Omega/\Box$) to a sharp Coulomb blockade    
(b: 8\,k$\Omega/\Box$). Samples with very high resistance       
show a backbending IVC (c: 40\,k$\Omega/\Box$).}                
\end{figure} 
Figure~\ref{resiivcfig} summarises the kinds of current-voltage
characteristics (IVC) we observed in 
the resistor samples. The data 
presented in fig.~\ref{resiivcfig} were taken
from wires on three different chips, anodised with different times and
final voltages, but with approximately equal dimensions. The
measurements were taken in four probe configuration at 
approximately (100\dots 200)\,mK in the
noncommercial dilution refrigerator, and in the absence of an externally
applied magnetic field.\par
For the most low resistive samples, like in trace (a), we found a remnant
of the supercurrent, visible as a region of reduced differential
resistance up to currents of about 2\,nA. This specific sample had a
sheet resistance at high bias or high temperature of approximately
1.5\,k$\Omega/\Box$. For samples anodised deeper, we observed an
increase of differential resistance around zero bias that we will from
now on refer to as Coulomb blockade (justification follows in section
\ref{transisection}, where we examine the samples in SET-like
geometry). \par
Trace (b) in figure~\ref{resiivcfig} is a typical example of a sharp
Coulomb blockade with a well-defined threshold voltage. This sample had a
sheet resistance of about 8\,k$\Omega/\Box$. In samples with very high
sheet resistance, we observed not only a sharp blockade, but even a
backbending of the current-voltage characteristic (trace c and inset in
fig.~\ref{resiivcfig}). The backbending alone might be related to
heating of the sample at the onset of current flow, where the relatively
high voltage leads to high power dissipation even at low currents. On
the other hand, this IVC shows a remarkable similarity with the
backbending IVC observed in arrays of ultrasmall Josephson junctions by
Geerligs 
et al. \cite{geerligs:89:jjaprl} and Chen 
et al. \cite{chen:92:ps}. \par
\begin{figure}[t]\begin{center} 
\epsfig{file=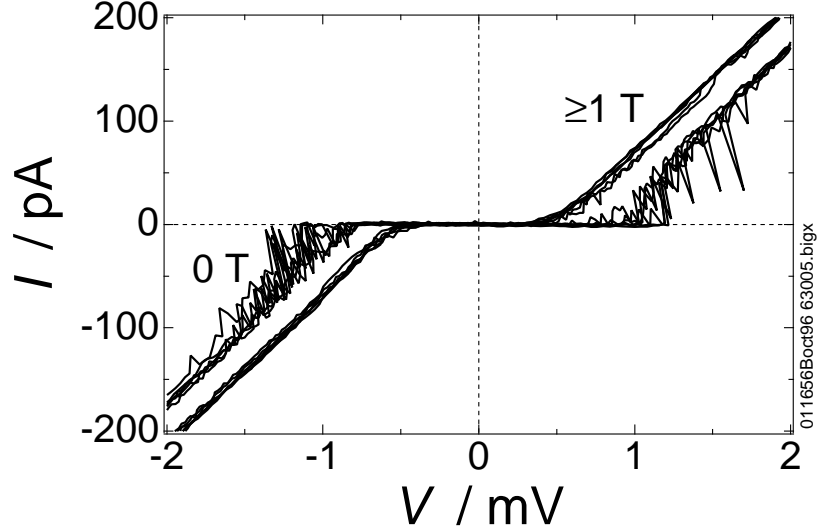,width=0.9\textwidth}      
\end{center}                                                    
\caption[Resistor IVC: magnetic field dependence; switching]{%
\label{magswitchfig}Effect of an external magnetic field on the IVC of a
high resistive sample. Offset and threshold voltages are reduced, and
the smoothing along the bias load line is smoothened out. Several traces
superimposed for both cases.}
\end{figure} 
Another feature that has been observed in more well-defined arrays of
superconducting junctions before is the switching of the current-voltage
characteristics between two or more envelopes near the
threshold. Figure~\ref{magswitchfig} gives an example of such
switching. The 
switching trace follows the load line defined by the biasing
resistors. The switching is random in time and bias voltage at which it
occurs in subsequent sweeps. We have observed that it suddenly settled
in single samples some minutes after start of a measurement. The
envelopes of the switching IVC seem to be reproducible, at least over a
time scale of minutes. In anodised niobium samples, we have observed
this switching both in the `old cryostat' that had no Thermocoax high
frequency filtering, and in the Oxford cryostat. These samples were not
measured in the Oxford cryostat before the Thermocoax filters were
installed, so that no comparison can be made in this respect. For more
information on switching in a chain of Josephson junctions and the
particular filtering and measuring setup, see the article by Haviland
et al. \cite{haviland:96:jvst}. \par
Another indication for a correlation between the switching of the IVC
and superconductivity is evident from fig.~\ref{magswitchfig}: here we
applied a magnetic field to quench superconductivity. The switching is
smoothend out, and the threshold voltage as well as the offset voltage
are reduced. This behaviour saturates  below 1.0\,T. Shown in
the figure are traces for external fields of (1.0 and 1.4)\,T, that
coincide within the measurement accuracy. The threshold voltage is
approximately reduced by a factor of two, which suggests that we might
be observing the Coulomb blockade of Cooper pair tunnelling evolving
into the Coulomb blockade for electrons.\par
\begin{figure}[t]\begin{center} 
\epsfig{file=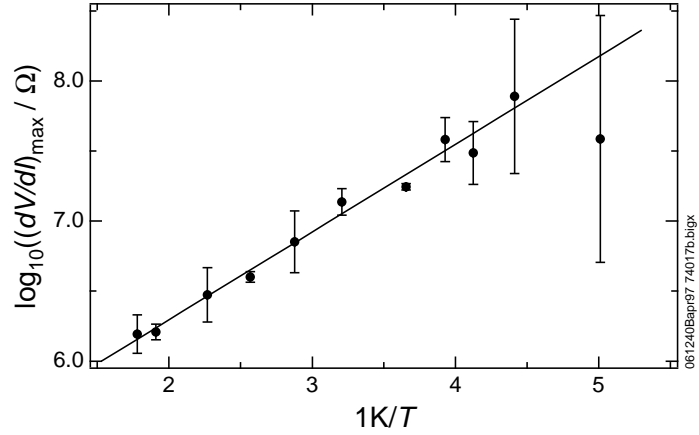,width=0.8\textwidth}      %
\end{center}                                                    %
\caption[Zero bias differential resistance: temperature dependence]%
{\label{zerobiasdiffresfig}
Zero bias differential resistance: temperature dependence. Determined
from numerical differentiation of measured $I$-$V$ curves.
No magnetic field applied.}
\end{figure} 
The temperature dependence of the resistance is the usual criterion for
classifying a material as insulating or having
a superconducting transition.
Our samples
have quite nonlinear IVC, so that assigning a `global' resistance is
rather futile. Instead, one often considers the differential resistance
at zero bias. A dedicated measurement would involve a careful biasing
scheme and a sensitive detection, possibly involving lock-in techniques
\cite{chen:94:thesis}. One can however, albeit with a substantial loss
of accuracy, extract some information about temperature dependence from
the measured IVC. The data plotted in figure~\ref{zerobiasdiffresfig}
are taken from numerically differentiated IVC. The centered value gives
the height of the differential resistivity peak around zero bias, and
the errors were estimated from the curvature of this peak at the
centre. For an intermediate temperature range, where the Coulomb
blockade is not fully developed, we see that the zero bias resistivity
follows an Arrhenius law,
\label{myarrh}
\begin{equation}
R_0(T)=R^\ast\cdot\exp\left(\frac{\DS E_a}{\DS k_B\,T}\right),
\end{equation}
suggesting a thermally activated behaviour \cite{chen:94:thesis}.\par
From figure~\ref{zerobiasdiffresfig}, we extract an activation energy
corrsponding to a temperature of about 0.6\,K or a voltage of
50\,$\mu$eV. This is the same order of magnitude as the voltage swing
and the temperature of disappearing voltage swing in our samples in
single electron transistor-like geometry, see~\ref{tempdepsubsection}.
More quantitative statements are complicated since this sample might
have been inhomogeneous on the length scale of micrometres due to a
skewed anodisation voltage profile. The average square resistance was
only about 4\,k$\Omega/\Box$
(total resistance 200\,k$\Omega$, dimensions
$10\times 0.2\,\mu$m$^2$), which should not have led to such a
pronounced blockade according to our other measurements, and which
confirms the suspicion of an inhomogeneous profile. The measurements
were done in the absence of an externally applied magnetic field.\par
\subsection{Measuring the Coulomb blockade: the offset voltage} 
\begin{figure}\begin{center} 
\epsfig{file=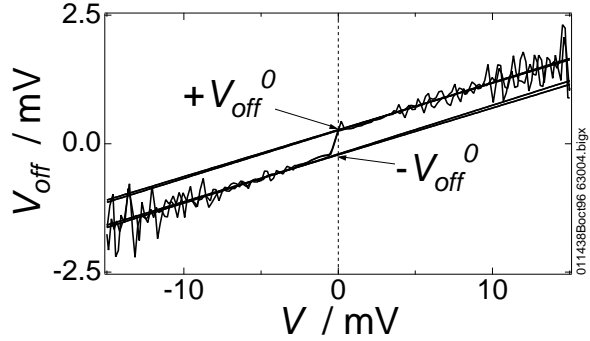,width=0.7\textwidth}      %
\end{center}                                                    %
\caption[Definition of $V_{\it off}^0$]{\label{voffzerodeffig}
Offset voltage. Extrapolation of                                %
$V_{\it off}(V)$, calculated from the tangent to the            %
$I(V)$ curve, to zero bias gives $V_{\it off}^0$, a measure     %
for the Coulomb blockade.}                                      %
\end{figure} 
Analysis of the IVC taken on our resistor samples show that they are
nonlinear over many decades in bias voltage. Therefore, the offset
voltage cannot simply be determined by extrapolating the tangent to the
IVC in whatever happens to be the edge of the particular measurement
interval. Instead, we follow the approach by  Wahlgren et al.
\cite{wahlgren:95:lic}, who has shown the value of an offset voltage
analysis for the understanding of the Coulomb blockade.\par
The offset voltage $V_{\it off}$
is treated as a quantity depending on the bias voltage $V$, and computed
numerically by extrapolating the tangent to the $I(V)$ curve to the
intersection with the voltage axis,
\begin{equation}
V_{\it off}(V)=V-I(V)\;
\left.\frac{dV^\prime}{dI}\right|_{V}.
\end{equation}
Figure~\ref{voffzerodeffig} shows the result of such a calculation. We
see that $V_{\it off}$ around zero bias is approximately a linear
function with an offset (so, it's not linear in the mathematical meaning
of the word). It is therefore possible to extrapolate the $V_{\it
off}(V)$ curve to zero bias, and the value at the intersection with the
$V_{\it off}$ axis, that we denote $V_{\it off}^0$, is well
defined. This value has been shown to be of particular interest in the
case of the Coulomb blockade in single junctions. 
$V_{\it off}$ gives the
limit for the blockade in the so-called `global rule' for low
environment impedances, where the whole electromagnetic environment
influences the Coulomb blockade \cite{wahlgren:95:prb}.\par
We performed four independent determinations of $V_{\it off}^0$ per IVC,
namely for negative and positive voltages, and for both directions of
the bias sweep.
\subsection{Onset of the CB: a superconductor-insulator transistion}
\begin{figure}\begin{center} 
\epsfig{file=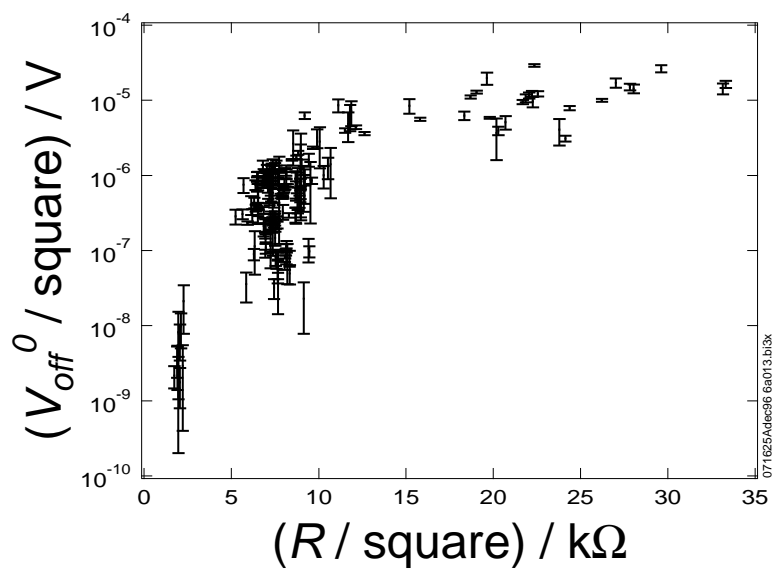,width=0.9\textwidth}      %
\end{center}                                                    %
\caption[Onset of Coulomb blockade in resistor samples]{
\label{cbonsetfig}Onset of Coulomb blockade (CB) in resistor    %
samples. Both $V_{\it off}^0$ and the sample resistance $R$     %
are normalised to the number of squares in the thin film        %
samples. The CB is clearly suppressed below                     %
6\,k$\Omega/\Box$. Error bars indicate the                      %
standard deviation of four                                      %
extrapolations for each set of data.}                           %
\end{figure} 
Figure~\ref{cbonsetfig} shows the results of such an offset voltage
analysis for a set of 187 strips
from four different chips  with length between (10 and
120)\,$\mu$m. The errors have been estimated as the standard deviation
of the four mentioned extrapolations per current-voltage
characteristic.\par
The correlation between extrapolated zero bias offset voltage and sample
resistance is obvious here where both quantities are normalised to the
number of squares in the film and plotted against each other. The
Coulomb blockade is appreciable for all samples with a sheet resistance
larger than 10\,k$\Omega/\Box$, and orders of magnitude less for samples
with less than 5\,k$\Omega/\Box$. In the intermediate region, there's a
lot of clutter due to the measurement errors, and possibly some
uncertainty in determining the sheet resistance. The sample width was
estimated based on scanning electron microscope inspection of samples
produced under similar conditions, and errors in square number
determination not indicated in fig.~\ref{cbonsetfig} might very well be
20\,\%. In any case, the Coulomb blockade sets in at
a value around 6\,k$\Omega/\Box$. It is obvious that we observed a
superconductor-insulator transition here, just in an unconventional
way. While `normally' the temperature dependence of the resistance is
considered, we looked at the magnitude of the Coulomb blockade. Of
course, this method per se cannot distinguish between superconducting
and normalconducting behaviour, so this information has to be derived
from the complete current-voltage characteristics.\par
It has been suggested from studies on ultrathin metal films
\cite{belitz:94:rmp,liu:93:sitprb} that there should be a universal
sheet resistance for the superconductor-insulator transition in such
films, and that it should be a fourth of the Klitzing resistance, namely
the so-called `superconducting quantum resistance'
\begin{equation}
R_Q=R_K/4=h/(4e^2)\approx 6.4\,\mbox{k}\Omega.
\end{equation}
Values that agree with this one
at least by order of magnitude, often
much better though, have not only been found in studies of quench-condensed
amorphous films 
\cite{jaeger:86:gallprb,haviland:89:2dscprl,jaeger:89:scprb}, but also
in regular arrays of ultrasmall Josephson junctions
\cite{geerligs:89:jjaprl,chen:92:ps,chen:95:prb}. \par
Our samples are obviously not regular arrays of Josephson junctions,
and not homegeneous, but granular. The average grain size is supposedly
a few nanometres, estimated
based on the futile attempts to see grains in the
scanning electron microscope and on literature data
\cite{halbritter:87:nbapa}. The superconducting coherence length, on the
other hand, should be around 40\,nm for 
good Nb \cite{auer:73:prb}. In
this case, the distinction between homogeneous and granular films should
vanish \cite{liu:93:sitprb}.
\section{Samples in SET-like geometry} 
\label{transisection}\label{setlikesection}
We put top gates on our resistor samples and found no modulation of the
IVC within the measurement 
accuracy. In the samples in single electron
transistor-like geometry, we were more successful.\par
This is not surprising, if we assume that the samples should behave like
arrays of very small Josephson junctions (Josephson junction arrays,
JJA). A modulation of the current-voltage characteristics 
with a gate voltage (`gate modulation') occurs when
the number of islands involved is small, especially when the transport
properties are dominated by a single island. Chandrasekhar et al.
\cite{chandrasekhar:91:prl,chandrasekhar:94:jltp} found charging effects
in rather short wires (0.75\,$\mu$m) of In$_2$O$_{3-x}$. This material
showed `the presence of large crystal grains'
\cite{chandrasekhar:91:prl}, and the authors concluded that
`only one or perhaps two segments (\dots) [were] present' 
in their samples.\par
On the other hand, the serration of the niobium metal's interface to the
(anodic) oxide film occurs on a length scale from 
below one nanometre up to a few nanometres
\cite{halbritter:87:nbapa}, so that all large grains in our thin films
should have been cracked, and the sample
should resemble an (irregular) array
of very many junctions.\par

\subsection{Current-voltage characteristics} 
Practically all 
SET-like samples showed a Coulomb blockade at millikelvin
temperatures. Often this Coulomb blockade manifested itself in a hardly
perceptible dip in the differential conductivity. In other samples, we
observed a rather sharp Coulomb blockade, and in several samples even at
liquid helium temperature, but in none of these cases did we manage to
modulate the current-voltage characteristics with an applied gate
voltage. Scanning microscope inspection of these samples often suggested
that there might have been a severe asymmetry between the two weak
links, either by damages to the bridge during evaporation or for example
by cracks in the bridge patched during evaporation. So it is quite
possible that we 
had often produced samples that behaved more like a single
junction than like a single electron transistor, and single junctions
show no gate modulation of the IVC.\par
\begin{figure}[t]\begin{center} 
\epsfig{file=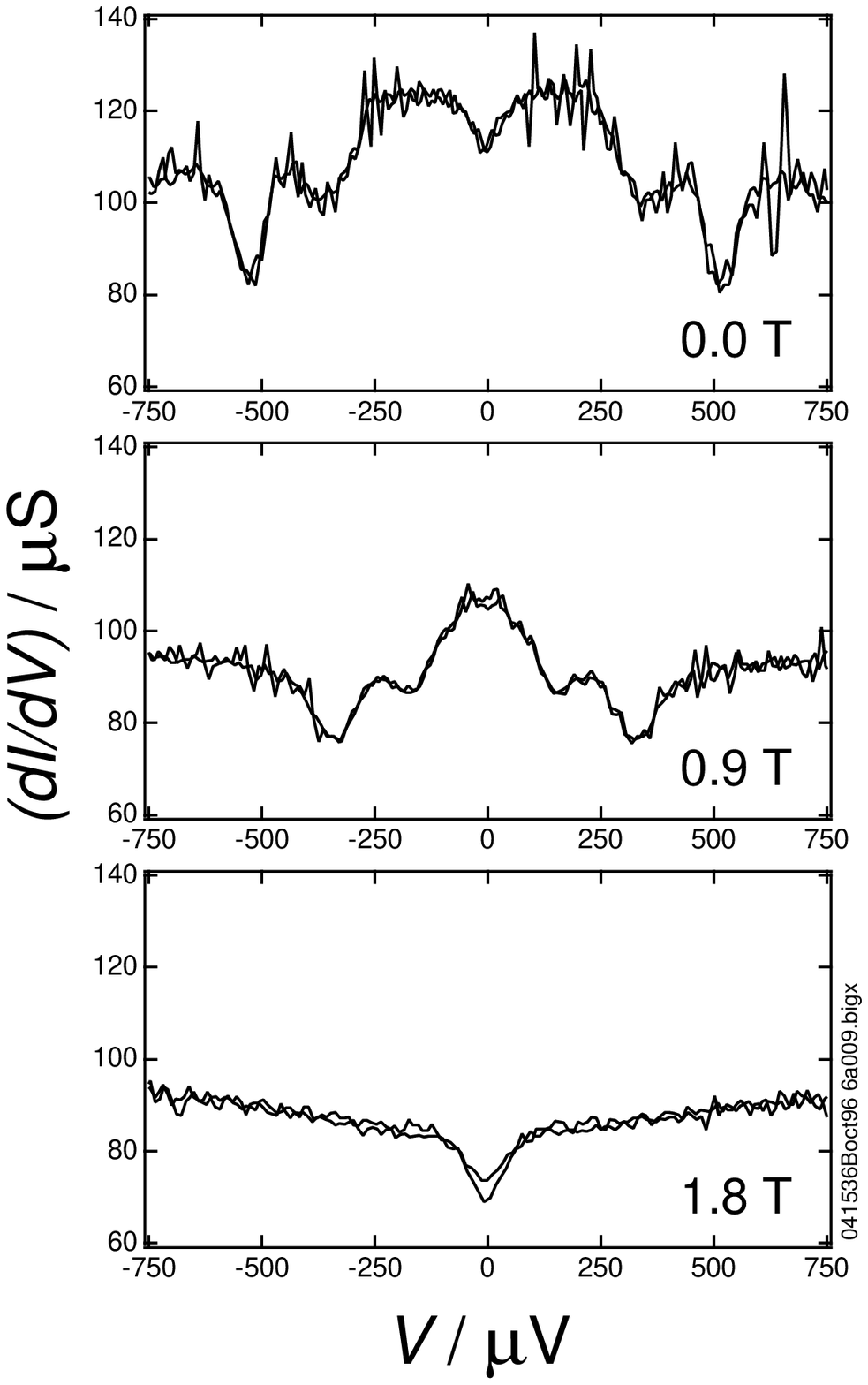,width=0.7\textwidth}      %
\end{center}                                                    %
\caption[Coulomb blockade in weak links sample]{
\label{weaklinkcbfig}Coulomb blockade in a sample with two      %
anodised weak links in SET geometry. As superconductivity       %
is squeezed by external magnetic fields, the off-zero-bias      %
differential conductance peaks disappear, and a Coulomb         %
blockade for electrons remains.}                                %
\end{figure}
\begin{figure}\begin{center} 
\epsfig{file=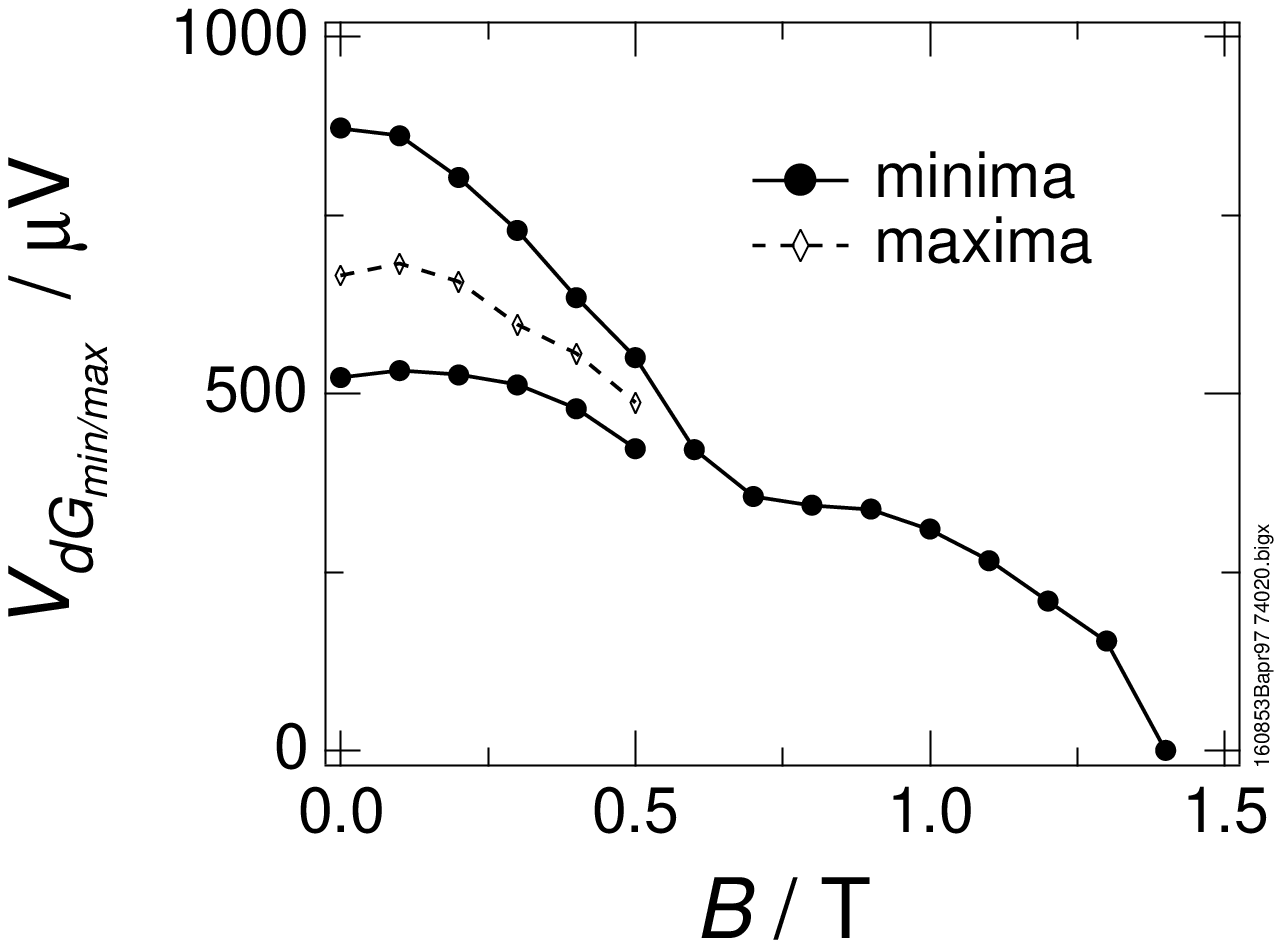,width=0.8\textwidth}
\end{center}
\caption[Magnetic field dependence of conductance extrema in weak links
sample]{\label{dipshiftfig}Magnetic field dependence of differential
conductance minima and maxima in figure~\ref{weaklinkcbfig}. Values
have been averaged from the positive and negative voltage
semiaxes. Lines are to guide the eye.}
\end{figure}

 
The samples whose IVC were susceptible to modulation by a gate voltage
had more complicated IVC, like in figure~\ref{weaklinkcbfig}. We plot
the differential conductivity here since the Coulomb blockade was weak,
and see a complicated system of conductivity peaks and dips around zero
bias. We  applied a magnetic field, and  these peaks and dips
moved toward zero bias. 
The differentiated IVC for vanishing, intermediate,
and high magnetic field are given in fig.~\ref{weaklinkcbfig}. The
highest (in bias voltage)
differential conductivity peaks and dips were easily identified, and in
figure~\ref{dipshiftfig} their location is plotted as a function of the
applied magnetic field.\par
At a field of 1.4\,T, all structures had disappeared except for a dip
in $dI/dV$ at zero bias. This dip obviously indicates the Coulomb
blockade of single electrons after the superconductivity has been
completely suppressed. The interpretation of the off-zero bias
structures is more complicated. The fact that they move continuously to
zero bias with magnetic field suggests that they are associated with the
superconducting energy gap. 
\par
\subsection{Control curves} 
\begin{figure}\begin{center} 
\epsfig{file=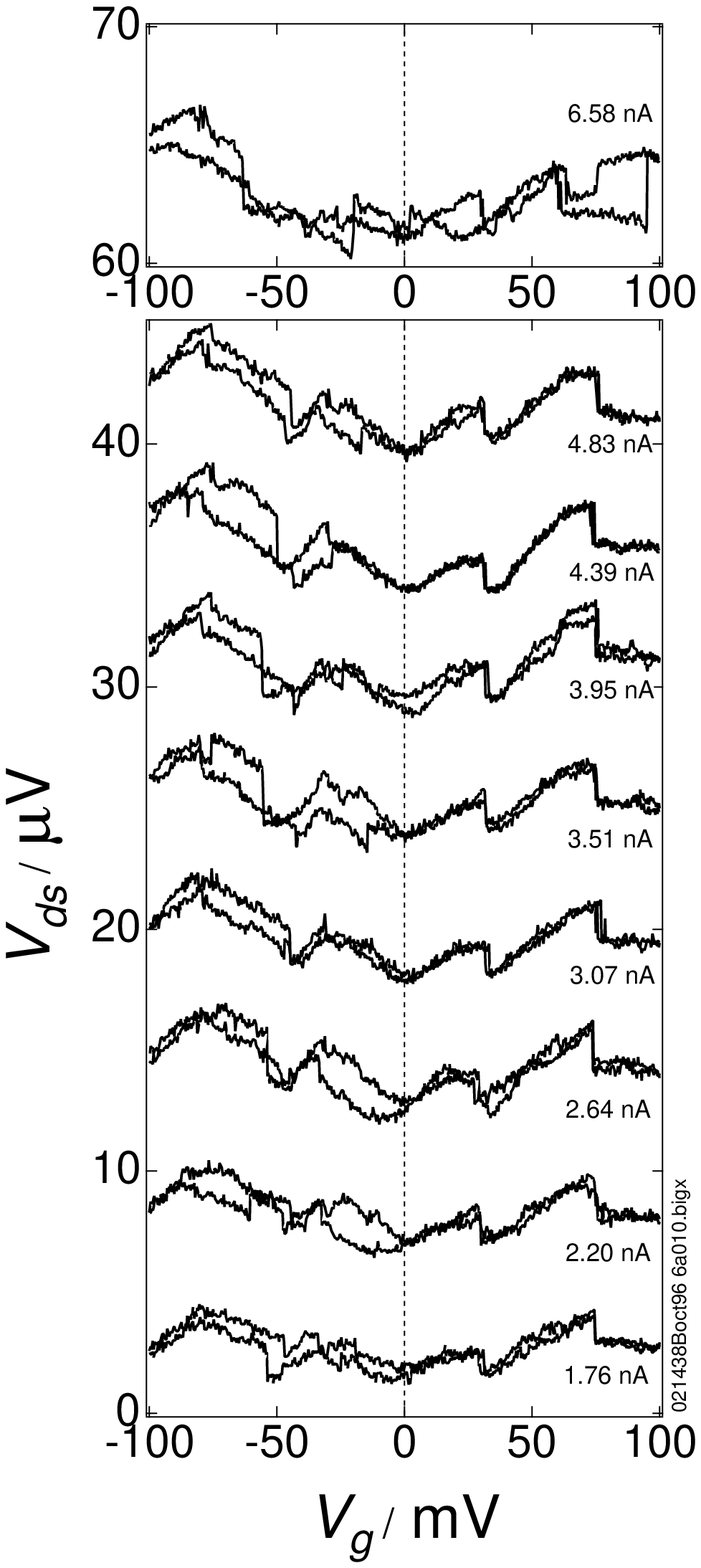,height=0.85\textheight}    %
\end{center}                                                    %
\caption[Control curves for weak links sample]{
\label{wealicontrfig}Control curves for a sample with two       %
anodised weak links in SET geometry. The drain-source voltage   %
$V_{\it ds}$ (at different current bias points) oscillates      %
as the voltage $V_g$ applied to a                               %
top gate is swept up and down.}                                 %
\end{figure} 
To see the response of the sample's current-voltage characteristic, we
biased it at a series of practically constant currents and swept the
gate voltage up and down with a frequency of about 8\,mHz. The voltage
between drain and source was registered in the usual way via the low
noise amplifiers in the cryostat top box. The gate voltage was
calculated from the voltage delivered from the gate voltage source and
the known division factors of the respective voltage dividers in use. The
design of the chip and the measurement wiring allowed to verify the
presence of the gate voltage on the chip through high frequency coaxial
leads.\par 
Figure~\ref{wealicontrfig} gives a series of $V_{\it ds}$-$V_g$
characteristics, that we refer to as `control curves', for the sample
whose differential current-voltage characteristics we examined in the
previous subsection. For this measurement, all superconducting effects
had been suppressed by applying an external magnetic field of 2\,T.\par
The causal influence of the gate voltage is obvious. The control curves
are far from the nice sine curves a single electron transistor would
give, but the correlation between the two sweep directions in gate
voltage is perceptible. The period of the voltage oscillations is of the
order of 50\,mV in gate voltage. If we treated the sample as a single
electron transistor, this would correspond to a total island capacitance
of the order of only 3\,aF. This is unrealistically low
if we assume that the charge of the middle island is
modulated.\par
Control curves with such a large periodicity are, however, typical for
systems of multiple tunnel junctions \cite{fulton:91:cbprl}. 
Periods of several V have been
observed in highly resistive superconducting microbridges
\cite{chen:93:mbieee,zhang:93:transieee}. A serial coupling of junctions
made by the step-edge cutoff technique \cite{altmeyer:95:apl} also gave
a large periodicity in gate voltage
\cite{altmeyer:96:sest}. Another system that has similar
transport properties are nanofabricated silicon wires
\cite{smith:97:siwirepreprint}. 
\subsection{Gate modulated IVC} 
\begin{figure}\begin{center} 
\epsfig{file=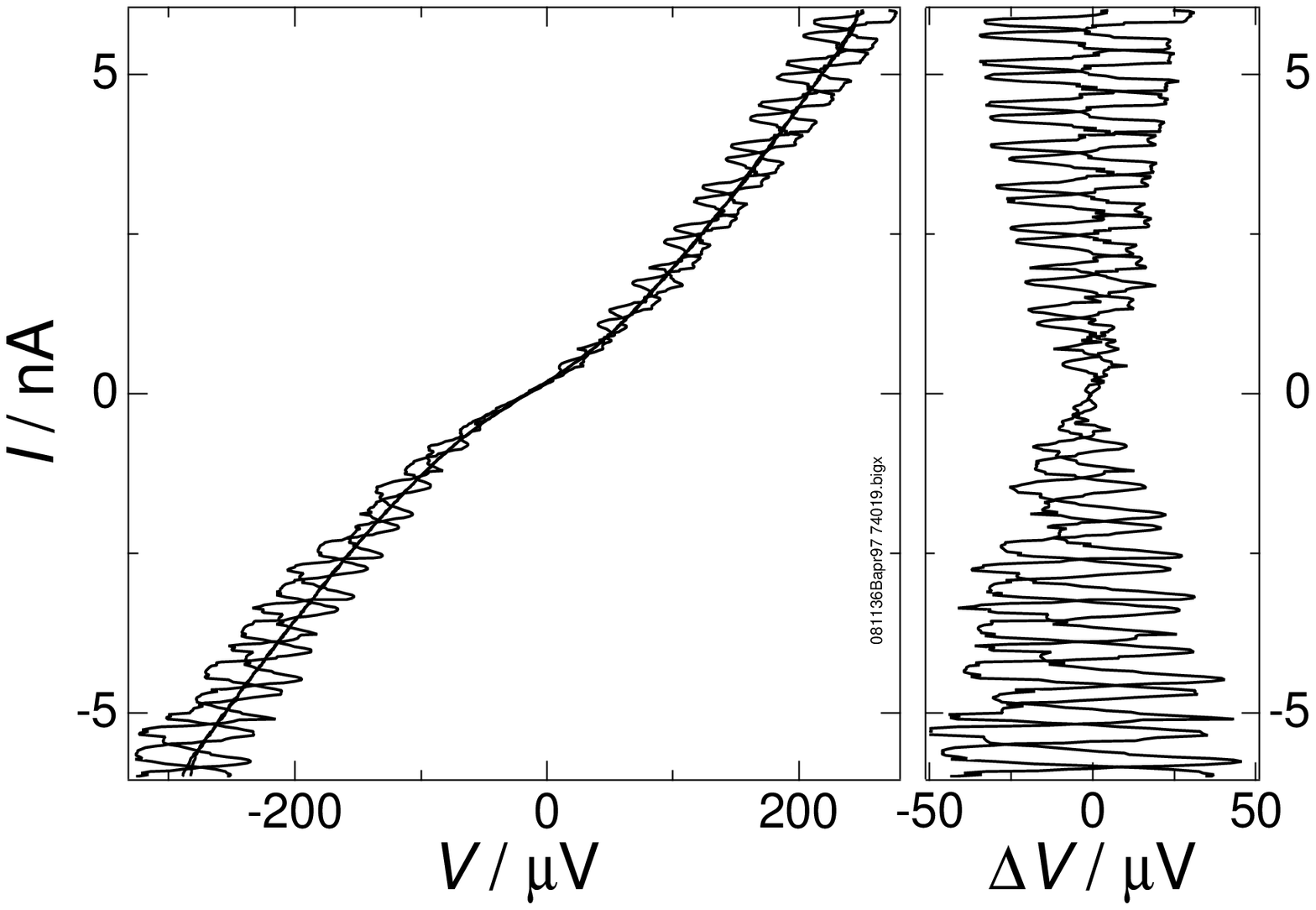,width=0.96\textwidth}     %
\end{center}                                                    %
\caption[Gate modulated $I$-$V$ characteristics
of weak links sample]{
\label{modulivcfig}Gate modulated $I$-$V$ characteristics. The gate
voltage was modulated with 40\,times the frequency of the bias sweep
(bias was ramped up and down once).
Left: resulting $I$-$V$ curve (wavy) and IVC in the absence of
a gate voltage (straight). Right: voltage swing $\Delta{}V$, defined as
the difference between modulated and unmodulated IVC.
$T\approx 50$\,mK, no external magnetic field.}
\end{figure} 
In one of the earliest papers on measurements on
single electron devices \cite{fulton:87:charprl}, Fulton 
and Dolan introduced a
technique of demonstrating the influence of a gate voltage on electronic
transport. If one sweeps the bias voltage slowly and simultaneously the
gate voltage with a higher frequency, the result is a trace zigzagging
around the unmodulated current-voltage characteristic. This technique
allows mapping out the entire modulation range with a single
measurement.\par
Figure~\ref{modulivcfig} gives the result of such a measurement on one
of the samples in single electron transistor-like geometry. Here the
bias was swept with a frequency of approximately 8\,mHz, and the gate
voltage with a frequency of 322.4\,mHz and an amplitude (peak-to-peak)
of 120\,mV. The measurement is part of a series where the gate amplitude
was up to 240\,mV, but at 120\,mV, the IVC were already modulated over
the maximum range.\par
In some samples, we observed an artefact created by capacitive pickup
rather than a modulation of the IVC. To ensure that the observation
plotted in fig.~\ref{modulivcfig} is not such an artefact, we checked
that the amplitude of the modulation did not depend on the frequency of
the gate voltage variation, at least not for frequencies of (80, 322.4
and 800)\,mHz. Secondly, we made sure that the deviation from the
unmodulated IVC (right panel in fig.~\ref{modulivcfig}) followed the
gate signal shape for sine, triangle, and square shape. In the case of
IVC variations
generated by capacitive pickup, the deviation followed the
time derivative of the gate signal, creating spikes in the case of
triangle and especially square shaped gate voltages.\par
Of course it would have been nice to map out the modulation range to
higher bias voltages, to see it decrease again,
presumably
(though not necessarily). Unfortunately, this particularly nice 
sample was destroyed
before such a measurement could be performed.\par

\subsection{Temperature dependence} 
\label{tempdepsubsection}
The temperature dependence of the gate modulation should give
information about the energy scale on which the Coulomb blockade
occurs. In the following, we present the results of the temperature
dependent measurements of the sample from figure~\ref{modulivcfig}. The
available data are sufficient 
to allow some comparison with single electron
transistors reported in the literature.\par
\begin{figure} 
\begin{center}
\epsfig{file=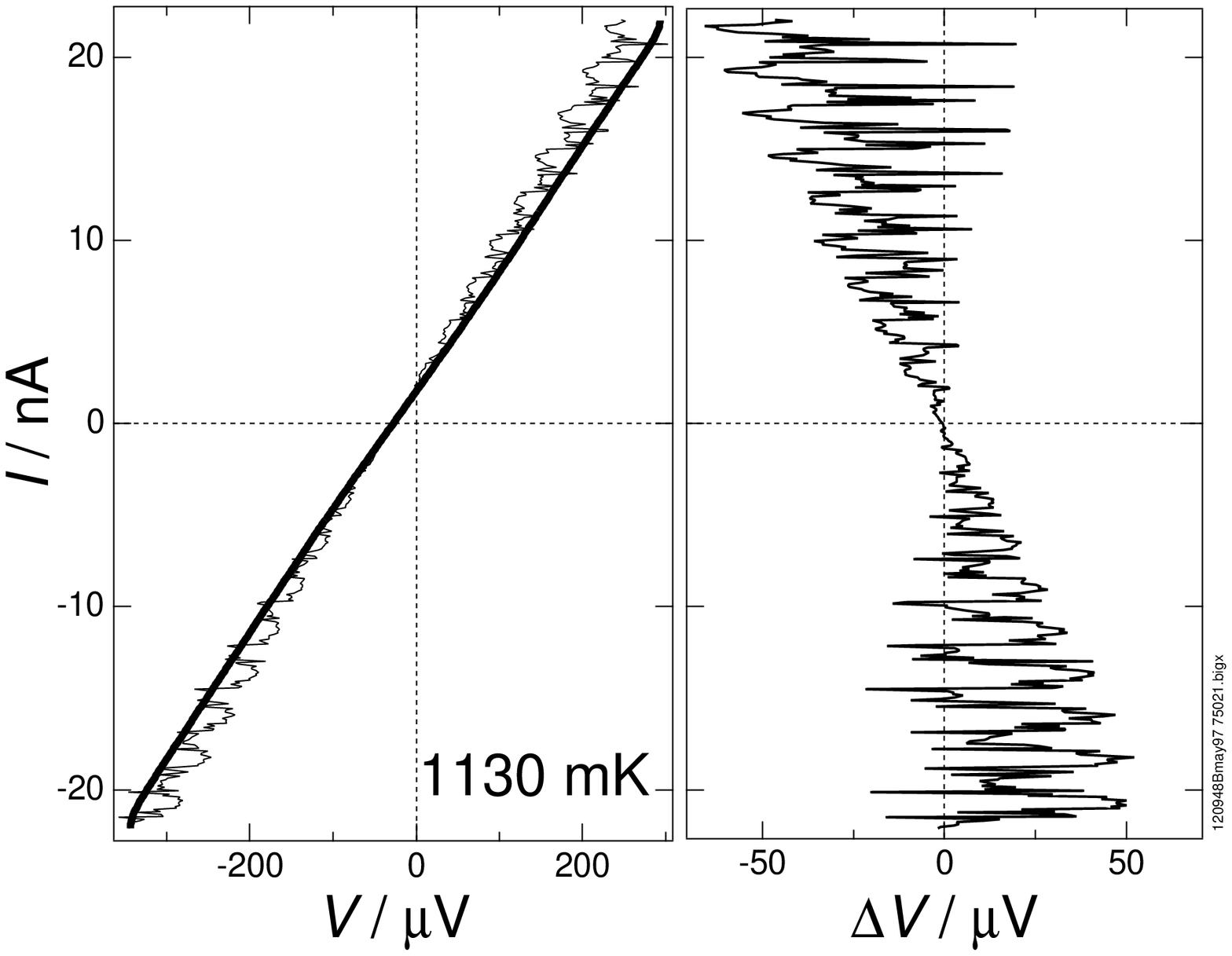,width=0.96\textwidth}
\caption[Gate modulation above 1\,K]%
{\label{onekelvinfig}
Modulation of the IVC with gate voltage (cf. fig.~\ref{modulivcfig}),
at a temperature of $(1130\pm 10)$\,mK. 
Only one direction of the bias ramping is
shown here. The thick trace in the left panel is an unmodulated IVC
taken a few minutes earlier, $\Delta{}V$ the difference to that IVC.
Gate voltage amplitude $V_{\it pp}=200$\,mV.}
\end{center}\end{figure} 
Figure~\ref{onekelvinfig} proves that the Coulomb blockade and its
modulation by the gate voltage persisted at temperatures above 1\,K. The
data were taken with the same method as that in figure~\ref{modulivcfig}.
The unmodulated current-voltage characteristic was taken a few minutes
before the modulated one. In the meantime, it had switched to another
trace, so that the difference between modulated and unmodulated IVC, the
voltage swing, is not symmetric around zero. Such a switching occurred
every few minutes, and is a well-known phenomenon in single electron
transistors.\par 
Most probable cause for the switching between several IVC is a change in
the configuration of the (random) background charge near the transistor
active structure. Background charge is an important contributor to noise
in single electron transistors. Oxides, like the barrier oxide or an
oxidised substrate, are a source of randomly fluctuating background
charges and thus of noise. In this respect, niobium resistors may be
rather disadvantageous, since they inevitably contain large amounts of
oxides.\par 

\begin{figure} 
\begin{center}
\epsfig{file=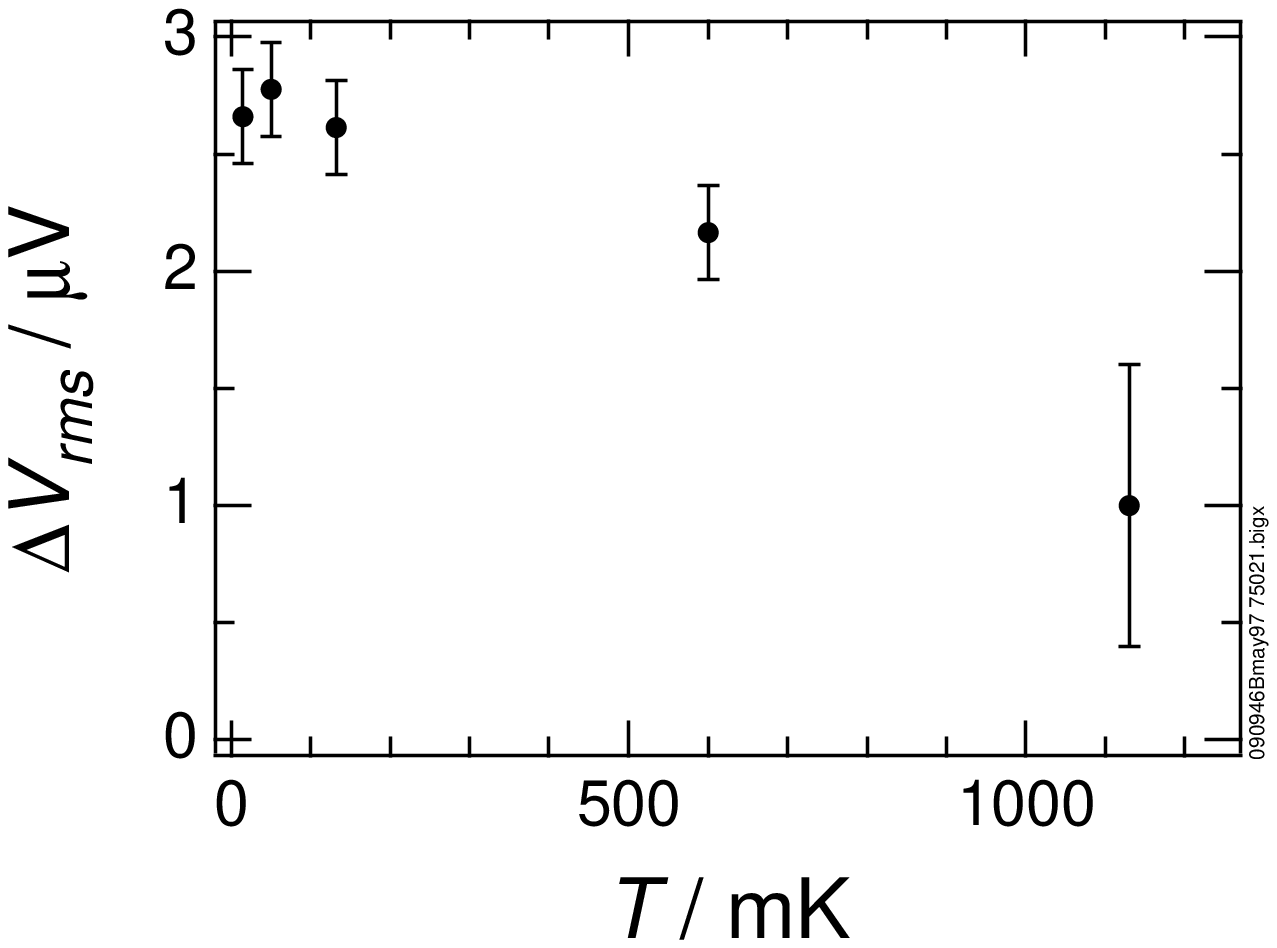,width=0.8\textwidth}
\caption[Temperature dependence of gate modulation]%
{\label{ampvstempfig}
Temperature dependence of the voltage swing for the sample from
figs.~\ref{modulivcfig} and \ref{onekelvinfig}. The first four values
have been obtained from an analysis of a calculation of the rms
amplitude in the bias region $(-7.5\dots 7.5)$\,nA for sine shaped gate
modulation, the last value by comparison of amplitudes at the edges of
this region for a square shaped modulation. The error bars indicate a
rough estimate.}
\end{center}
\end{figure} 
The temperature dependence of the voltage swing is shown in
figure~\ref{ampvstempfig}. For the low temperature values up to 600\,mK,
the swing $\Delta{}V$ was analysed in the bias region between $-7.5$\,nA
and 7.5\,nA, and the root-mean-square value of the amplitude is plotted
together with some estimate of the uncertainty, versus the
temperature. Errors in the temperature measurement were negligible on
this scale. The high temperature at 1130\,mK in
figure~\ref{ampvstempfig} was determined by comparison of the amplitudes
of the voltage swing at $\pm 7.5$\,nA in two measurements with a square
shaped gate modulation and normalised to the low temperature
amplitude.\par
As expected, the amplitude of the modulation decreases with
temperature. A simple extrapolation of the few data points in
figure~\ref{ampvstempfig} gives a temperature value between 2\,K and
3\,K where the voltage swing vanishes and which we denote as
$T^\ast$. The peak-to-peak maximum voltage swing 
found for this sample at low
temperature was at least 100\,mK; it might have been somewhat bigger, if
one had been able to measure on this sample at higher bias. This
corresponds to a temperature of about 1.2\,K, or about one half 
$T^\ast$.\par
For single electron transistors made by angular evaporation, Wahlgren
\cite{wahlgren:94:xjobb} found an approximate relation
\begin{equation}
\label{idealseteq}
e\Delta{}V_{\it max}\approx 4 k_B T^\ast.
\end{equation}
This relation seems to hold even for the very small single electron
transistors made by Nakamura et al. \cite{nakamura:96:set100k} that had
a $T^\ast$ of the order of 100\,K. In both these cases, the junction
resistances were considerably above the quantum resistance $R_K$.\par
Our sample considered above, on the other hand, had a total resistance
of less than $R_K$. In such low-resistive samples, cotunnelling plays an
important role, that is the `simultaneous' tunnelling through a virtual
state between two tunnel barriers. Cotunnelling could account for the
reduction of the voltage swing compared to the `ideal' single electron
transistor described by (\ref{idealseteq}).
\chapter{Conclusion}
\section{Status quo}
We have successfully adapted and developed methods for the fabrication
of high-ohmic low-capacitance resistors. By means of anodic oxidation,
these resistors can be trimmed with a spread in resistance values of
less than 10\% for a value of a few hundred k$\Omega$ on the length of a
few tens of micrometres. There is no need to monitor and anodise each
resistor independently; however, one monitored resistor per anodisation
process is required since the resistance depends sensitively on film
thickness, and on anodisation voltage and time in a way far too complex
to allow a prediction based trimming.\par
We have found that our nanofabricated resistors showed nonlinear
current-voltage characteristics, with a transistion from superconducting
behaviour to a Coulomb blockade with increasing sheet resistance. 
A purely ohmic behaviour is practically impossible to achieve, though
one might suppress the supercurrent with a magnetic field. But even in
this case, a (weak) Coulomb blockade remained. The Coulomb blockade grew
by orders of magnitude when the sheet resistance of the films exceeded a
value that seems to agree with the quantum resistance for Cooper pairs
$R_K/4\approx 6.45$\,k$\Omega$.\par
We have fabricated a device based on two weak links in a niobium thin
film strip, defined by shadow evaporation and anodisation, in a geometry
resembling that of a single electron transistor. These samples could be
equipped with an overlapping gate thanks to the insulating properties of
the anodic oxide film. By applying a voltage to the gate, the
conductance of the device could be modulated.\par
The behaviour resembled closely that of a single electron transistor,
but the shape of the control curves suggests that we were in fact dealing
with multiple tunnel junctions in 100\,nm wide structures. A
further reduction of size seems hard to achieve; however, changes in
design might alleviate the problem of asymmetry between the junction
arrays on both sides of the transistor island.\par
For low resistive, transistor-like samples, a suppression of the voltage
modulation compared to single electron transistors was found that can be
explained by cotunnelling. Due to the multiple junction nature of the
weak links, there is a tradeoff between a tendency to a
more gate controllable
blockade at lower resistance (because fewer junctions are involved), and
the weakening of this blockade due to cotunnelling (for the same
reason). 

\section{Directions for future research}
The superconductor-insulator transition should be investigated with
samples in different (wider) geometry to corroborate the found
dependence of the Coulomb blockade on the sheet resistance. 
Temperature and magnetic field dependent data 
on the superconductor-insulator transition
are needed.
With the
four layer resist technique presented above, however, results would have
been compromised by the grainy surface contamination occurring on large
exposed surfaces. First tests with three layer resist (PMMA-Ge-PMGI)
are very promising for the fabrication of structures from 100\,nm to
many micrometres without surface contamination. Therefore, this
technique would also enable the fabrication of well-defined junction
arrays for comparative studies. Prerequisite is the availability of good
e-beam evaporated niobium, and test runs of a new niobium evaporation
system (April 1997) indicate that we can now make Nb thin films with
transition temperature close to the bulk value in 100\,nm thick
films.\par
It remains to use anodisation fabricated niobium thin film resistors for
the purpose they were originally intended for, namely as biasing
resistors for ultrasmall tunnel junctions, in an attempt to learn more
about the interplay of the Coulomb blockade and the electromagnetic
environment.\par 
The single electron transistor-like samples should be made with an
improved fabrication process, and characterised systematically by
magnetic field dependent and especially temperature dependent
measurements. 
\cleardoublepage\pagestyle{headings}
\addcontentsline{toc}{chapter}{Bibliography}

%
\begin{appendix}
\chapter{Symbols and notation}
\label{symbolsapp}
\topcaption{Meaning, SI unit and numerical value 
    \cite{cohen:93:const} (if applicable) of %
    symbols used in this report} 
\tablehead{\hline Symbol\vphantom{{\Large (}} & 
     meaning (numerical value) & 
     SI unit \\\hline}
\tabletail{\hline}
\begin{supertabular}{|p{0.10\textwidth}|p{0.55\textwidth}|p{0.2\textwidth}|}
$\Delta$ \vphantom{{\Large (}} &
   superconducting energy gap &
   m$^2$\,kg\,s$^{-2}$ \\
$\Delta{}V$ \vphantom{{\Large (}} &
   voltage swing (modulation) &
   m$^2$\,kg\,s$^{-3}$\,A$^{-1}$ \\
$\gamma$ \vphantom{{\Large (}} &
   phase difference over Josephson junction &
   1 \\
$\varepsilon$ \vphantom{{\Large (}} &
   relative dielectric permittivity &
   1 \\
$\varepsilon_0$ \vphantom{{\Large (}} &
   dielectric permittivity of vacuum ($8.854\dots\cdot 10^{-12}$)&
   m$^{-3}$\,kg$^{-1}$\,s$^{4}$\,A$^2$  \\
$\lambda$ \vphantom{{\Large (}} &
   London penetration depth &
   m \\
$\mu_0$ \vphantom{{\Large (}} &
   permeability of vacuum ($1.256\dots\cdot 10^{-6}$) &
   m\,kg\,s$^{-2}$\,A$^{-2}$ \\
$\xi$ \vphantom{{\Large (}} &
   coherence length &
   m \\
$\varrho$ \vphantom{{\Large (}} &
   density &
   m$^{-3}$\,kg \\
$\phi$ \vphantom{{\Large (}} &
   phase of multiparticle wavefunction &
   1 \\
$A$ \vphantom{{\Large (}} &
   area &
   m$^2$ \\
$B$ \vphantom{{\Large (}} &
   magnetic flux density &
   kg\,s$^{-2}$\,A$^{-1}$ \\
$C$ \vphantom{{\Large (}} &
   capacitance &
   m$^{-2}$\,kg$^{-1}$\,s$^{4}$\,A$^{2}$ \\
$d$ \vphantom{{\Large (}} & 
   thickness &
   m \\
$e$ \vphantom{{\Large (}} & 
   elementary charge ($1.602\dots\cdot10^{-19}$) &
   s\,A \\
$E$ \vphantom{{\Large (}} &
   energy &
   m$^2$\,kg\,s$^{-2}$ \\
$E_a$ \vphantom{{\Large (}} &
   activation energy &
   m$^2$\,kg\,s$^{-2}$ \\
$E_C$ \vphantom{{\Large (}} &
   characteristic charging energy $e^2/(2C)$&
   m$^2$\,kg\,s$^{-2}$ \\
$E_{\it ch}$ \vphantom{{\Large (}} &
   charging energy &
   m$^2$\,kg\,s$^{-2}$ \\
$E_J$ \vphantom{{\Large (}} &
   Josephson (coupling) energy &
   m$^2$\,kg\,s$^{-2}$ \\
$G$ \vphantom{{\Large (}} &
   conductivity &
   m$^{-2}$\,kg$^{-1}$\,s$^{3}$\,A$^{2}$ \\
$H$ \vphantom{{\Large (}} &
   magnetic field &
   m$^{-1}$\,A \\
$h$ \vphantom{{\Large (}} &
   Planck's constant ($6.626\dots\cdot 10^{-34}$) &
   m$^2$\,kg\,s$^{-1}$ \\
$\hbar$ \vphantom{{\Large (}} &
   Planck's constant divided by $2\pi$ ($1.054\dots\cdot 10^{-34}$) &
   m$^2$\,kg\,s$^{-1}$ \\
$I$ \vphantom{{\Large (}} &
   current &
   A \\
$I_c$ \vphantom{{\Large (}} &
   critical current &
   A \\
$I_{\it c0}$ \vphantom{{\Large (}} &
   maximum critical current &
   A \\
$I_s$ \vphantom{{\Large (}} &
   supercurrent &
   A \\
$k_B$ \vphantom{{\Large (}} &
   Boltzmann constant ($1.380\dots\cdot 10^{-23}$) &
   m$^2$\,kg\,s$^{-2}$\,K$^{-1}$ \\
$l$ \vphantom{{\Large (}} &
   length &
   m \\
$Q$ \vphantom{{\Large (}} & 
   charge  &
   s\,A \\
$R$ \vphantom{{\Large (}} & 
   resistance &
   m$^2$\,kg\,s$^{-3}$\,A$^{-2}$ \\
$R_K$ \vphantom{{\Large (}} & 
   quantum resistance &
   m$^2$\,kg\,s$^{-3}$\,A$^{-2}$ \\
$R_{\mbox{\footnotesize K-90}}$ \vphantom{{\Large (}} & 
   Klitzing resistance (25812.807) &
   m$^2$\,kg\,s$^{-3}$\,A$^{-2}$ \\
$R_Q$ \vphantom{{\Large (}} & 
   `quantum resistance for pairs' $R_K/4$ &
   m$^2$\,kg\,s$^{-3}$\,A$^{-2}$ \\
$R_T$ \vphantom{{\Large (}} & 
   tunnelling resistance &
   m$^2$\,kg\,s$^{-3}$\,A$^{-2}$ \\
$t$ \vphantom{{\Large (}} &
   time &
   s \\
$T$ \vphantom{{\Large (}} & 
   temperature &
   K \\
$T^\ast$ \vphantom{{\Large (}} & 
   temperature at which voltage swing vanishes &
   K \\
$T_c$ \vphantom{{\Large (}} & 
   critical temperature &
   K \\
$V$ \vphantom{{\Large (}} & 
   voltage &
   m$^2$\,kg\,s$^{-3}$\,A$^{-1}$ \\
$V_{\it off}$ \vphantom{{\Large (}} & 
   offset voltage &
   m$^2$\,kg\,s$^{-3}$\,A$^{-1}$ \\
$V_{\it off}^0$ \vphantom{{\Large (}} & 
   (extrapolated) zero bias offset voltage &
   m$^2$\,kg\,s$^{-3}$\,A$^{-1}$ \\
$x$ \vphantom{{\Large (}} &
   spatial coordinate &
   m \\
\end{supertabular}
\chapter{Glossary and abbreviations}
\label{glossary}
\begin{description}
\item[\AA{}ngstr\"om (\AA{}):] Outdated unit of
   length. 1\,\AA{}$=10^{-10}$\,m. 
\item[ACME:] Anodization controlled miniaturization enhancement
   \cite{nakamura:96:acme}
\item[AF:] Anodic film (more general than AOF, may incorporate
   inclusions from the electrolyte)
\item[Ammonium pentaborate:] (NH$_4$)B$_5$O$_8\cdot x$H$_2$O
   \cite{apbdictionaryentry}, where $x$ 
   indicates the amount of crystal water. 
   If undefined, we assume $x\approx 4$ (APB tetrahydrate).
\item[AOF:] Anodic oxide film
\item[APB:] Ammonium pentaborate
\item[ASCII:] American Standard Code for Information Interchange
\item[CAD:] Computer Aided Design
\item[Cb:] Chemical symbol for Columbium
\item[CB:] Coulomb blockade
\item[CBCPT:] Coulomb blockade of Cooper pair tunnelling
\item[CHET:] Charging effect transistor
\item[Chip marks:] Alignment marks (JEOL EBL system) used for highest
   precision alignment. Three chip marks have to be situated within one
   field. In the work described here, wafer marks were used instead.
\item[Columbium (Cb):] old name for Niobium (Nb), used in the
   Angloamerican language space until about 1950 and in the American
   metallurgical community even later.
\item[Contrast:] Degree to which the physicochemical properties
   exploited in resist development differ in exposed areas compared to
   unexposed areas
\item[CP:] Copolymer P(MMA-MAA)
\item[DAQ:] Data acquisition
\item[DMM:] Digital multimeter
\item[DMS:] Dilute magnetic semiconductor
\item[DXF:] Drawing Exchange Format
\item[DVM:] Digital voltmeter
\item[ECU:] European currency unit
\item[EBL:] Electron beam lithography
\item[EOS:] Electron optical system
\item[Field:] Area that can be written by the EBL without the need of
   moving the stage, $80\times 80\,\mu$m$^2$ in highest resolution mode. At
   the edges of the fields, stitching error occurs, so fine structures
   should not cross field boundaries.
\item[Forming:] The process of growing an anodic [oxide] film
\item[Gauss (G):] Hopelessly outdated unit of magnetic flux density in
   one of the various CGS systems. Corresponds to (but is not equal to)
   0.1\,mT.
\item[GDS-II:] Stream format, a.\,k.\,a. Calma Stream. The industry
   standard for lithographic pattern data.
\item[GPIB:] General purpose interface bus (IEEE-488)
\item[IBE:] Ion beam etching (milling)
\item[Inch:] Outdated unit of length. 1\,in$=25.4$\,mm
\item[IUPAC:] International Union of Pure and Applied Chemistry
\item[IVC:] Current-voltage characteristic
\item[JEOL:] Japanase manufacturer of electron optical research
   instrumentation 
\item[JJ:] Josephson junction
\item[JJA:] Josephson junction array
\item[Jobdeck file (JDF):] Text file in the JEOL EXPRESS
   system that describes which patterns (chips) are to be exposed and
   where they are to be positioned relative to each other, 
   and that contains the definition of the shot modulation
   and the alignment (mark detection) parameters.
   It also points to a calibration sequence of the electron optical
   system that will be carried out at the beginning of the exposure
   and eventually during the exposure.
\item[Microfabrication:] Fabrication of devices with typical linear
   dimensions below 1\,$\mu$m
\item[M-IT:] Metal-insulator transition
\item[ML:] Monolayer
\item[Nanofabrication:] The art and science of producing non-random
   structures with typical linear dimensions less than 100\,nm
\item[Negative resist:] Resist that is removed during development where
   it has \emph{not} been exposed. Example is SAL\,601 (e-beam resist).
\item[PMGI:] Poly(dimethyl glutarimide), a positive e-beam and deep UV
   resist 
\item[PMMA:] Polymethylmethacrylate, a positive e-beam resist
\item[P(MMA-MAA):] Copolymer
   poly(methylmethacrylate-methacrylic acid), a positive e-beam resist,
   more sensitive than PMMA
\item[Positive resist:] Resist that is removed during development where
   it has been exposed. Examples are PMMA (e-beam resist) or S-1813
   (photoresist). 
\item[PROXECCO:] A commercial computer programme for proximity
   correction \cite{eisenmann:93:proxecco}.
\item[Proximity correction:] Increasing the exposure dose for narrow
   and/or isolated features to compensate for the proximity effect.
\item[Proximity effect:] Additional exposure of pixels with many
   neighbouring exposed pixels due to scattering of the electron beam in
   the resist and substrate and to secondary electrons.
\item[QPT:] Quantum phase transition
\item[RIE:] Reactive ion etching
\item[Rotation:] Angular misorientation of the  sample relative to the
   sample holder and 
   consequently the whole electron beam lithography machine.
   Rotation has to be compensated by the EOS, increasing inaccuracies
   (stitching error) and pattern distortions. A limit on 
   allowed rotation is
   set in the internal configuration files of the JEOL system.
\item[RRR:] Residual resistance ratio, between the resistances at room
   temperature and just above the resistive transition or at 4.2\,K; a
   measure for the quality often used for Nb. 
\item[S-1813:] A positive photoresist
\item[SAL\,101:] A developer for PMGI
\item[SAL\,601:] A negative e-beam resist
\item[Schedule file (SDF):] Text file in the JEOL EXPRESS
   system that describes where the arrangement of patterns defined in the
   jobdeck file is to be placed relative to the machine and what the
   reference dose for the shot modulation is.
   It also contains information on hardware settings and definitions
   for the alignment mark detection.
\item[Selectivity:] Ratio of the solubilities of different resists
   exposed simultaneously, important for the resolution in processes
   involving two layer resist systems
\item[SEM:] Scanning electron microscope
\item[Sensitivity:] Reciprocal of the irradiation dose required to
   produce the physicochemical modifications in a resist needed for
   development
\item[SET:] Single electron tunnelling, alt. single electron
   (tunnelling) transistor
\item[Shadow evaporation technique:] also known as Dolan technique,
   Niemeyer-Dolan technique, nonvertical evaporation technique etc.
   A method of forming very small overlap
   junctions in the shadowed area underneath a suspended bridge on the
   substrate. Self-aligning, involves only one lithography
   step. Introduced by Niemeyer \cite{niemeyer:74:mitt}, in its present
   form with resist mask by Dolan \cite{dolan:77:masks}.
\item[Shot modulation:] JEOL-specific implementation of handling
   the assignment of doses to pattern parts (to compensate the
   proximity effect). Each primitive is assigned a shot rank 
   (an integer number) that
   corresponds to a certain dose enhancement factor
   (a floating point number). This assignment
   is called the shot modulation.
\item[S-IT:] Superconductor-insulator transition
\item[SNAP:] Selective niobium anodization process
   \cite{kroger:81:anodiapl} 
\item[SnL:] Swedish Nanometre Laboratory, G\"oteborg.
\item[Stitching error:] Misalignment of parts of the electron
   beam exposed pattern at the boundaries of fields and subfields.
   Stitching error increases with sample rotation.
\item[Subfield:] Area that can be written by the EBL without switching
   digital-to-analogue converters, $10\times 10\,\mu$m$^2$ in highest
   resolution mode. At subfield boundaries, slight stitching 
   error occurs, so the
   finest nanostructures should not cross them.
\item[Tear-off technique:] A special form of angular evaporation
   technique where some material is deposited on resist sidewalls and
   removed during liftoff. Requires good control over the undercut and
   the evaporation angles.
\item[TEM:] Transmission electron microscopy
\item[UHV:] Ultra high vacuum, below $10^{-6}$\,Pa
\item[Vector scan:] EBL mode where the beam is swept only over the areas
   that are to be exposed, as opposed to raster scan, where it is swept
   over the whole sample and simply blanked from non-exposure
   areas. Requires faster electron optics and makes systems more
   expensive, but can save a lot of exposure time.
\item[VTB:] Variable thickness bridge
\item[Wafer marks:] Alignment marks (JEOL EBL system) that can be placed
   almost anywhere on the sample. Of course, precision of alignment
   improves when the marks are as close to the writing area as possible.
\item[ZEP\,520:] A positive e-beam resist
\end{description}
\chapter{Recipes}
\label{recipes}
All recipes assume that reactive ion etching (RIE) is done in a
Plasmatherm Batchtop 70 with a seven inch electrode (area
248\,cm$^2$), an electrode distance of 60\,mm and a working
frequency of 13.56\,MHz.
\par
The contact printer 
operates in the wavelength range 
(320\dots 420)\,nm$^2$.\par
Electron beam lithography was done with a JEOL JBX 5D-II system
with CeB$_6$  cathode.
\section{Photomask making}
\begin{enumerate}
\item Rinse a Cr mask with deionised tap water.
\item Ash the surface with oxygen RIE, pressure 33\,Pa,
flow 36\,$\mu$mol/s, rf power 50\,W, time 30\,s.
\item Spin Microposit Primer.
\item Spin Shipley SAL-601 at 4000\,rpm, giving a thickness of about
800\,nm.
\item Preexposure bake for 20\,min at 90\,$^\circ$C
   in an oven.
\item E-beam expose in the JEOL JBX 5D-II. Design dose
10\,$\mu$C/cm$^2$, acceleration voltage 50\,kV, fourth lens
   (working distance 39\,mm),
   third aperture
   (diameter 300\,$\mu$m), current 5\,nA.
\item Postexposure bake for 20\,min at 110\,$^\circ$C
   in an oven.
\item Develop in Microposit MF322 for about 6\,min, inspect in the
microscope.
\item Ash the surface in the RIE (see above) and immediately 
   thereafter
\item etch in Balzers No.\,4 chromium etch 
(composition: 200\,g cerium ammonium nitrate, 35\,mL 98\% acetic acid,
filled with deionised water to 1000\,mL).
\item Remove the resist by stripping with RIE. Process gas oxygen, pressure
66\,Pa, flow 7\,$\mu$mol/s, rf power 250\,W, time 120\,s.
\end{enumerate}

\section{Gold pad photolithography (carrier chips)}
\label{padrecipe}
\begin{enumerate}
\item Strip the surface 
of an oxidised two inch Si wafer with RIE.  Process gas oxygen, pressure
66\,Pa, flow 7\,$\mu$mol/s, rf power 250\,W, time 120\,s.
\item Spin Shipley S-1813 at 5500\,rpm, giving a thickness of about
1000\,nm.
\item Bake for 7:30\,min at 110\,$^\circ$C
   on a hotplate.
\item Expose for 12\,s at an intensity of 10\,mW/cm$^2$, correspondingly
longer or shorter for different intensities.
\item Develop in a 1:1 mixture (by volume) of Microposit Developer and
de\-ion\-ised water for 60\,s, rinse thoroughly with deionised water from
the tap. \textbf{Or:}
\item Develop in pure MF\,322 developer for 15\,s and rinse.
\item Ash the surface with oxygen RIE, pressure 33\,Pa,
flow 36\,$\mu$mol/s, rf power 50\,W, time 30\,s. Immediately 
thereafter
\item evaporate 20\,nm of Ni$_{0.6}$Cr$_{0.4}$ 
at 0.1\,nm/s and
\item 80\,nm Au at 0.2\,nm/s.
\item Liftoff in slightly warmed acetone.
\item Presaw from the back to a depth of about 50\,$\mu$m. When cutting
alignment edges from the front side, try to preserve a C$_4$
symmetric circumference shape of the wafer; this facilitates later
resist preparation.
\end{enumerate}

\section{Four layer resist preparation}
\begin{enumerate}
\item Ash the surface 
of a wafer with gold chip patterns
with RIE. Process gas oxygen, pressure 33\,Pa,
flow 36\,$\mu$mol/s, rf power 50\,W, time 30\,s.
\item Spin 350k PMMA (1.8\,\%, in xylene) at 2500\,rpm to a thickness
of about 50\,nm.
\item Bake for 12\,min at 170\,$^\circ$C
   on a hotplate.
\item Spin Shipley S-1813, diluted 1:1 by volume with Shipley
P-Thinner, at 3000\,rpm, giving a thickness of about 200\,nm.
\item Bake for 12\,min at 160\,$^\circ$C
   on a hotplate.
\item Evaporate 20\,nm Ge at 0.2\,nm/s.
\item Spin 350k PMMA (1.8\,\%, in xylene) at 2500\,rpm to a thickness
of about 50\,nm.
\item Bake for 10\,min at 150\,$^\circ$C
   on a hotplate.
\item Break into suitable chip sets for further handling.
\end{enumerate}

\section{Four layer resist exposure}
\label{4lrexposurerecipe}
Acceleration voltage 50\,kV,
   first aperture
   (diameter 60\,$\mu$m), fifth lens
   (working distance 14\,mm), current 20\,pA for the fine patterns
   (1\,nA for the coarser leads).
   Area doses
\begin{itemize}
\item 1120\,$\mu$C/cm$^2$ for 20\,nm wide lines.
\item 400\,$\mu$C/cm$^2$ 
   for 100\,nm wide lines.
\item 280\,$\mu$C/cm$^2$ 
   for all wider lines and areas.
\end{itemize}

\section{Four layer resist proximity correction}
\label{proxparapp}
(for PROXECCO:) double Gaussian with $\alpha=0.006\,\mu$m,
$\beta=6\,\mu$m and $\eta=0.5$. The low $\eta$ is due to the Ge layer
that absorbs a large fraction of the backscattered electrons. Number of
doses 32, output quality fine, physical fracturing.

\section{Four layer resist processing}
\label{flpprcapp}
\begin{enumerate}
\item Expose in the EBL machine (see \ref{4lrexposurerecipe}).
\item Develop in a mixture of 10 volume parts isopropanole and 1
   volume part deionised water for 60\,s under ultrasonic
   excitation.
\item Reactive ion etching: pattern transfer to the Ge mask.
   Process gas CF$_4$, pressure 1.3\,Pa, flow 
   7.5\,$\mu$mol/s,
   rf power 14\,W, time 120\,s.
\item RIE of the support layers. Process gas O$_2$, pressure
   13\,Pa, flow 
   15\,$\mu$mol/s, rf power 20\,W, time 15\,min.
\item Evaporate Nb with e-gun heating. Deposition rate about
   0.5\,nm/s.
\item Liftoff in slightly warmed acetone, spraying chip centres
   directly with a syringe.
\end{enumerate}

\section{Anodisation window mask}
\begin{enumerate}
\item Spin 950k PMMA (8\,\%, in 
chlorobenzene) at 5000 rpm, giving a thickness
of about 1.8\,$\mu$m.
\item Bake for 12\,min at 170\,$^\circ$C
   on a hotplate.
\item E-beam expose with an  area dose of
280\,$\mu$C/cm$^2$. Acceleration voltage 50\,kV,
   first aperture
   (diameter 60\,$\mu$m), fifth lens
   (working distance 14\,mm), current 1\,nA.
\item Develop in a mixture of 10 volume parts isopropanole and 1
volume part deionised water under ultrasonic excitation for 8\,min.
\end{enumerate}

\section{Electrolyte for Nb anodisation}
Downscaled from the recipe of Joynson \cite{joynson:67:anodi}:
8.3\,g ammonium pentaborate, 60\,mL ethylene glycole and 40\,mL
distilled water to be stirred and heated to about 100$^\circ$\,C.
The solution has to be regenerated by heating and stirring before
using since the ammonium pentaborate precipitates.

\section{Two layer resist for high resolution EBL}
\label{twolayerrecipe}
\begin{enumerate}
\item Spin copolymer (6\,\%, in 2-ethoxy-ethanole) at 5000\,rpm, to a
thickness of about 140\,nm.
\item Bake for 5\,min at 170\,$^\circ$C
   on a hotplate.
\item Spin NANO-PMMA (2\,\%, in anisole) at 5000\,rpm, giving a film
thickness of about 50\,nm.
\item Bake for 5\,min at 170\,$^\circ$C
   on a hotplate.
\item E-beam expose.
\item Develop in a mixture of 10 volume parts isopropanole and 1
volume part deionised water under ultrasonic excitation for 50\,s.
\end{enumerate}

\section{Ti dot patterns (etch mask for IBE) on II-VI
semiconductors}
\label{tidotmaskrecipe}
\begin{enumerate}
\item Prepare two layer resist (see \ref{twolayerrecipe}).
\item E-beam expose with an acceleration voltage of 50\,kV, first
aperture (diameter 60\,$\mu$m), fifth lens (working distance 14\,mm),
current 1\,nA. Dose 
\begin{itemize}
\item 700\,$\mu$C/cm$^2$ for $50\times 50$\,nm$^2$ sqaures on a
   250\,nm periodic square lattice,
\item 220\,$\mu$C/cm$^2$ for $200\times 200$\,nm$^2$ sqaures on a
   400\,nm periodic square lattice.
\end{itemize}
\item Develop (see \ref{twolayerrecipe}).
\item Evaporate 30\,nm Ti.
\item Liftoff in warm acetone. Apply jet from a syringe needle,
   this may take a while.
\end{enumerate}

\section{Ar$^+$ ion beam milling of II-VI semiconductor quantum
dots}
\begin{enumerate}
\item Make Ti dot mask (see \ref{tidotmaskrecipe}).
\item Mill for 20\,min under normal incidence. Acceleration voltage
200\,V, current density 0.16\,mA/cm$^2$. Etching rate under these
conditions is approximately 20\,nm per minute for 
Cd$_{1-x}$Mn$_x$Te.
\item Remove the Ti in 10\,\% HF (a few seconds), rinse with water and
isopropanole.
\end{enumerate}

\section{Chemical etching of II-VI semiconductor quantum dots}
\begin{enumerate}
\item Dissolve one drop of Br$_2$ in 10\,mL ethylene glycol.
\item Etch for half a minute.
\item Check under the SEM.
\item Repeat etch and check until satisfied.
\end{enumerate}
\end{appendix}
\end{document}